\documentclass[aps, twocolumn,groupedaddress,nofootinbib,nobalancelastpage,nobibnotes]{revtex4}
\pdfoutput=1
\usepackage{amsmath,amsfonts,amssymb,mathrsfs,graphicx,color}
\usepackage[squaren]{SIunits}
\usepackage{verbatim}
\usepackage{enumerate}
\usepackage{graphicx}
\usepackage{ulem}
\usepackage[hyperfootnotes=false]{hyperref}
\usepackage{xcolor}
\usepackage{slashed}
\usepackage[utf8]{inputenc}
\usepackage[mathscr]{euscript}

\newcommand{\ie}{{\it i.e.}}

\newcommand{\eg}{{\it e.g.}}

\newcommand{\eq}{Eq.}
\newcommand{\eqs}{Eqs.}

\newcommand{\fig}{Fig.}

\newcommand{\Ref}{Ref.}
\newcommand{\Refs}{Refs.}
\newcommand{\Sec}{Section}

\newcommand{\App}{Appendix}

\newcommand{\Tab}{Tab.}

\newcommand{\equ}[1]{\eq~(\ref{equ:#1})}
\newcommand{\figu}[1]{\fig~\ref{fig:#1}}

\newcommand{\bi}{\begin{itemize}}
\newcommand{\ei}{\end{itemize}}
\newcommand{\ra}{\rightarrow}
\newcommand{\lra}{\longrightarrow}

\newcommand{\rmd}{{\rm d}}
\newcommand{\pgamma}{{\rm p}\gamma}

\newcommand{\nue}{\nu_e}
\newcommand{\nuebar}{{\bar \nu}_e}
\newcommand{\numu}{\nu_\mu}
\newcommand{\numubar}{{\bar \nu}_\mu}
\newcommand{\nutau}{\nu_\tau}
\newcommand{\nutaubar}{{\bar\nu}_\tau}
\newcommand{\nua}{\nu_\alpha}
\newcommand{\nub}{\nu_\beta}

\newcommand{\rarr}{\rightarrow}
\newcommand{\ba}{\begin{array}}
\newcommand{\ea}{\end{array}}

\newcommand{\SP}{{\mathscr P}}

\begin{document}

\title{Astrophysical Neutrino Production Diagnostics with the Glashow Resonance}

\date{\today}

\author{Daniel Biehl}
\affiliation{DESY, Platanenallee 6, 15738 Zeuthen, Germany}

\author{Anatoli Fedynitch}
\affiliation{DESY, Platanenallee 6, 15738 Zeuthen, Germany}

\author{Andrea Palladino}
\affiliation{Department of Astroparticle Physics, Gran Sasso Science Institute, Via Francesco Crispi 7, 67100 L’Aquila, Italy}

\author{Tom J. Weiler}
\affiliation{Department of Physics \& Astronomy, Vanderbilt University, Nashville TN 37235, USA}

\author{Walter Winter}
\affiliation{~DESY, Platanenallee 6, 15738 Zeuthen, Germany~~}

\begin{abstract}
We study the Glashow resonance $\nuebar + e^- \rightarrow W^- \rightarrow$ hadrons at 6.3 PeV as diagnostic of the production processes of ultra-high energy neutrinos. The focus lies on describing the physics of neutrino production from pion decay as accurate as possible by including the kinematics of weak decays and Monte Carlo simulations of pp and ${\rm p}\gamma$ interactions. 
We discuss optically thick (to photohadronic interactions) sources, sources of cosmic ray nuclei, and muon damped sources. Even in the proposed upgrade IceCube-Gen2, a discrimination of scenarios such as pp versus ${\rm p}\gamma$ is extremely challenging under realistic assumptions. Nonetheless, the Glashow resonance can serve as a smoking gun signature of neutrino production from photohadronic (${\rm A}\gamma$) interactions of heavier nuclei, as the expected Glashow event rate exceeds that of pp interactions. We finally quantify the exposures for which the non-observation of Glashow events exerts pressure on certain scenarios.
\end{abstract}

\maketitle
\tableofcontents

\section{Introduction}
\label{sec:intro}

The evidence for high-energy neutrinos from space in the IceCube experiment~\cite{Aartsen:2013jdh} has opened a new window into the Universe from the perspective of a different messenger. While the origin of these neutrinos is still debated, it seems unlikely that a single class of conventional sources, such as Active Galactic Nuclei (AGN) blazars~\cite{Glusenkamp:2015jca,KowalskiNeutrino2016}, Gamma-Ray Bursts (GRBs)~\cite{Abbasi:2012zw,Aartsen:2014aqy}, or starburst galaxies~\cite{Murase:2013rfa,Bechtol:2015uqb} can power the observed diffuse flux on their own, unless the sources are hidden, such as collimated jets inside stars~\cite{Murase:2013ffa}. It therefore is also important to study conceptual arguments, related to the spectral shape (such as the power law index), see \eg\ \Refs~\cite{Aartsen:2015zva,Aartsen:2015knd,Aartsen:2016xlq} for different data analyses, and the flavor composition~\cite{Aartsen:2015knd} of astrophysical neutrinos. Other currently debated issues are the possible contribution from a (possibly softer) galactic component especially in the Southern hemisphere \cite{Palladino:2016zoe,PV2}, and a possible hardening of the spectrum at high energies~\cite{KowalskiNeutrino2016}. 

Since the current information is inclusive, new generic diagnostic tools are needed to identify the source class. One example are damped muons in the pion decay chain~\cite{dampedMU,Kashti:2005qa}, which is typically found for strong magnetic fields $B \gtrsim 10^3 \, \mathrm{G}$~\cite{Hummer:2010ai,Baerwald:2011ee}. Another example, which we focus on in this study, is the Glashow resonance event rate as an indicator for the electron antineutrino contribution to the total flux. 
The Glashow resonance commonly refers to resonant scattering $\nuebar + e^- \to W^- \to {\rm anything}$ 
at $E_\nu \simeq 6.3~{\rm PeV}$~\cite{Glashow:1960zz,Berezinsky:1977sf}.
Note that this process is unique for $\nuebar$, as the 
 interaction rate of $\nue$, $\numu$, $\nutau$, $\numubar$, $\nutaubar$ with electrons is negligible compared to
interactions with nucleons.
The Glashow resonance has been widely studied from the viewpoint of the predicted event rate and discrimination power for pp versus ${\rm p}\gamma$ interactions~\cite{Anchordoqui:2004eb,Hummer:2010ai,Bhattacharya:2011qu,Xing:2011zm,Bhattacharya:2012fh,Barger:2012mz,Barger:2014iua,Palladino:2015uoa,Shoemaker:2015qul,Anchordoqui:2016ewn,Laha},  
 which are generic source classes indicative for \eg\ starburst galaxies (pp) versus AGNs/GRBs (${\rm p}\gamma$). While the discrimination between pp and ${\rm p}\gamma$ sources seems unlikely in the current IceCube experiment, there are plans for a volume upgrade IceCube-Gen2~\cite{Aartsen:2014njl} and a large experiment in sea water, called KM3NeT-ARCA~\cite{Adrian-Martinez:2016fdl}, which may be capable to perform this measurement.
 
In this study, we evaluate realistic subtleties in the cosmic production of neutrino flavor fluxes
to determine the power of Glashow events at  6.3~PeV to discriminate among the most popular source models. We describe  
the current (simplified) paradigms for neutrino production, propagation, and (Glashow) detection in \Sec~\ref{sec:flavor}. In \Sec~\ref{sec:exp}, we include the kinematics of the secondary decays, and we test the production paradigms using hadronic Monte Carlo event generators for pp and ${\rm p}\gamma$ interactions. \Sec~\ref{sec:cond} discusses optically thick sources, sources of heavier nuclei, and damped muon sources. 
We finally interpret the non-observation of Glashow events in \Sec~\ref{sec:non} and conclude in  \Sec~\ref{sec:conc}. Details on spectrum weighted moments of neutrinos and pions, the effect of $K$ mesons, and the cross check with IceCube effective areas can be found in the Appendix.

\section{Flavor and neutrino-antineutrino composition}
\label{sec:flavor}

Here we discuss the flavor and neutrino-antineutrino composition of the astrophysical neutrinos under somewhat idealized conditions, which correspond to the state-of-the-art in a large portion of the current literature, and we present our methods. We split this discussion into source, propagation, detection, and  methods.

\subsection{Composition at the Source}
\label{sec:AstroModels}

   The signal for
$\nuebar$ at the Glashow resonance,  normalized to the total
$\nu + \bar \nu$ flux, can be used to differentiate among the main 
primary mechanisms for neutrino-producing interactions in 
optically thin sources of cosmic rays~\cite{Anchordoqui:2004eb}. 
For example, in idealized ${\rm p}\gamma$ interactions, the process
\begin{equation}
p+\gamma \rightarrow \Delta^+ \rightarrow \begin{cases} \pi^+ + n &\text{ 1/3 of all cases} \\ \pi^0 + p &\text{ 2/3 of all cases}\end{cases}
\label{equ:delta}
\end{equation}
will lead,  after pion decay
\begin{eqnarray}
\pi^+ & \rightarrow & \mu^+ + \numu \, ,\nonumber \\
& & \mu^+ \rightarrow e^+ + \nue + \numubar \, , \label{equ:piplusdec}  
\end{eqnarray}
to a neutrino population with $N_{\numu}=N_{\bar\numu} =
N_{\nue} \gg N_{\bar\nue}$ (referred to as the {\it ``$\pi^+$ mode''}). For the pp collision mechanism, a nearly
isotopically neutral mix of pions is expected from isospin 
invariance
\begin{equation}
{\rm p+p} \rightarrow \begin{cases} \pi^+ + \text{anything} &\text{ 1/3 of all cases} \\ 
\pi^- + \text{anything} &\text{ 1/3 of all cases} \\ 
\pi^0 +  \text{anything} &\text{ 1/3 of all cases} \end{cases}  \, .
\label{equ:idealpp}
\end{equation}
As a result,  a neutrino
population with the ratio $2N_{\numu}=2N_{\bar\numu} =
N_{\nue}=N_{\bar\nue}$ is expected (referred to as the {\it ``$\pi^\pm$ mode''}). Note that these estimates do not include the kinematics of pion and muon decays.

Other popular source models include muon damping, which means that the muons lose energy by synchrotron or adiabatic losses faster than they can decay. These modes are  ${\rm pp}\rarr \pi^\pm$~pairs $\rarr \numu,\ \numubar$ only, referred to as the {\it ``damped $\mu^\pm$ mode''}, and  $\pgamma\rarr\pi^+ \rarr \numu$ only, referred to as the {\it ``damped $\mu^+$ mode''}. We show the flavor composition at the source for the popular production modes in the second column of \Tab~\ref{table:nuebar}, and we defer the discussion of the muon damping in greater detail to a later section.

Another production scenario is neutron decay into a pure $\nuebar$ beam~\cite{Anchordoqui:2003vc}, which  is outside the notion of the pion decay scenario discussed here.
The large value of Earthly $\nuebar$ in this case implies a large rate of Glashow events, that are not observed.  
Although neutrons are produced in all hadronic interactions (pp and ${\rm p}\gamma$), the anti-neutrino from its decay receives kinematically a very small energy fraction. These energies are typically much smaller than the Glashow energy, see \Ref~\cite{Hummer:2010ai}. In the following we will not consider pure neutron decay any further. 

\begin{table*}[hbt]
\caption
{Neutrino flavor composition at the source and fraction of $\nuebar$ in total neutrino flux at Earth (corresponding to the relative strength of the Glashow resonance)  for four  popular astrophysical source models. Neutrinos and antineutrinos are shown separately, when they differ. Here the  TBM approximation for flavor mixing is used, and the  kinematics of pion and muon decays are not included.
}
\label{table:nuebar}
\centering
\begin{tabular}{|c|c|c|c|c|c|}
 \hline 
 &\multicolumn{2}{c|}{\ Source flavor composition\ \ }  & \multicolumn{2}{c|}{\ Earthly flavor composition\ \ }   &\ Earthly $\nuebar$ fraction $\xi^f_{\nuebar}$  \\
 & \multicolumn{2}{c|}{\ $(\phi_e:\phi_\mu:\phi_\tau)$ \ \ } & \multicolumn{2}{c|}{\ $(\phi^f_e:\phi^f_\mu:\phi^f_\tau)$ \ \ } &  in cosmic neutrino flux\ \ \\
\hline
 ${\rm pp}\rarr\pi^\pm$~pairs\ \ & \multicolumn{2}{c|}{(1:2:0)} & \multicolumn{2}{c|}{(1:1:1)} & $9/54=17\%$   \\ 
\ \ w/ damped $\mu^\pm$  & \multicolumn{2}{c|}{(0:1:0)} & \multicolumn{2}{c|}{(4:7:7)} &  $6/54=11\%$  \\ 
\hline
 & $\nu$ & $\bar \nu$ &  $\nu$ & $\bar \nu$ & \\ 
\hline
\ $\pgamma\rarr\pi^+$ only\ \ & \ (1:1:0)\ \ & (0:1:0) & (14:11:11) & (4:7:7) & $4/54=7.4\%$ \\ 
\ \ \ w/ damped $\mu^+$ & \ (0:1:0)\ \ & (0:0:0) & (4:7:7) & (0:0:0) &  0  \\ 
\hline
\end{tabular}
\end{table*}

All in all, the approximate results presented in this section must be treated as suggestive. It is the purpose of this paper to confront idealized fluxes with a more realistic modeling. 
For example, it has been shown in \Refs~\cite{Hummer:2010ai,Winter:2012xq}, that multi-pion contributions can ameliorate the Glashow event rate difference between these two models. 
For certain source parameters, the ``contamination'' from multi-pion processes can be large~\cite{Hummer:2010vx}.
In addition, muon damping at the sources is possible; it may be complete damping or partial (incomplete).
Finally, there is the effect of kaon production and decay on source neutrino flavor ratios;
we show below that the effect is small in the energy range 
of interest~\cite{Hummer:2010ai,Winter:2012xq}.

\subsection{Propagation Effects}
\label{sec:propagation}

Here we discuss the propagation from source to detector from flavor mixing, using the idealized assumptions for the neutrino production described above, and tri-bimaximal values for the neutrino mixing angles; we refer to the mixing effects with the prefix ``TBM''. 
Note that in \Sec~\ref{sec:exp} and later, we will however include the kinematics of muon and pion decays to define our {\em ideal} pp and {\em ideal} ${\rm p}\gamma$ sources. 

We follow \Refs~\cite{Fu:2012zr,Fu:2014isa} here and take the tri-bimaximal mixing model~\cite{Harrison:2002kp,Harrison:2002et} as a good approximation for the flavor mixing. 
Then, the evolution $\nua\rarr\nub$, with $\alpha$ and $\beta$ any elements of the three-flavor set
$\{e,\mu,\tau\}$, 
is described in terms of the PMNS matrix $U$ by the symmetric propagation matrix $\SP$.
Let the general flavor composition at the source be denoted by $(\phi_e:\phi_\mu:\phi_\tau)$ and at the detector by $(\phi^f_e:\phi^f_\mu:\phi^f_\tau)$, representing the initial and final neutrino fluxes, where this composition may describe the sum over neutrinos and antineutrinos, or neutrinos and antineutrinos separately.
The conversion from initial flavor basis to propagating mass basis and back to flavor basis at Earth is affected by the transition amplitude
$A_{\alpha\ra\beta}=\sum_j U_{\alpha j} e^{-iE_{\nu_j} L} U_{j \beta}$. 
Over large astronomical distances, the oscillating interference terms average out, 
and one obtains a (3-flavor$\times$3-flavor) probability matrix 
\begin{equation}
\label{eq:Prop}
\SP(\alpha\ra\beta)=\sum_j | U_{\alpha j}|^2 | U_{j \beta} |^2\,,\ \ {\rm relating\ }\vec{\phi}^f = \SP \vec{\phi}\,.
\end{equation}
In the TBM model, the probability elements are given by
\begin{equation}
\label{eq:Ptheta13}
\SP_{\rm TBM}(\alpha\ra\beta)= 
\frac{1}{18}
\left(
\ba{rrr}
 10 & 4 & 4 \\
  4 & 7 & 7 \\
  4 & 7 & 7 \\
\ea
\right)\,.
 \end{equation}
The identical nature of the 2nd and 3rd rows/columns of $\SP_{\rm TBM}$ reflects the $\mu$-$\tau$ symmetry,
which results from the TBM assumption of a maximal $\theta_{32}$ and zero $\theta_{13}$.
The symmetry means that muon and tau neutrinos arrive at Earth with the same equilibrated probability regardless of 
their original ratio, \ie, $\phi^f_\mu=\phi^f_\tau$.
The Earthly flavor composition for the popular production scenarios can be found in the third column of \Tab~\ref{table:nuebar}.

Note that  tau neutrinos are not expected at the source in the pion production scenarios, \ie, $\phi_\tau = 0$, because the charged partner $\tau^\pm$ has a mass  much larger than the pion mass. 
As a result, we obtain~\cite{Fu:2012zr,Fu:2014isa}
\begin{equation}
\label{equ:PPT2by2}
\left(
\begin{array}{c}
\phi^f_e \\
\phi^f_{\not e}
\end{array}
\right)
 = \SP^{(2\times2)}_{\rm TBM}
\left(
 \begin{array}{c}
 \phi_e \\
 \phi_\mu
 \end{array}
\right)
 \,,
\SP^{(2\times2)}_{\rm TBM} = \frac{1}{9}
\left(
\begin{array}{ccc}
 5 & 2 \\
 4 & 7    
\end{array}
\right) 
\end{equation}
and the Earthly flux $\phi^f_{\not e}$ evenly distributed between $\numu$ and $\nutau$.

The rate of resonant Glashow events is directly proportional to the $\nuebar$ content in the neutrino flux at the detector, whereas the rate of neighboring continuum events is proportional to the whole neutrino flux.
It is therefore useful to  introduce the quantity $\xi_{\bar \nu_\ell}$ as the fraction of  antineutrinos of flavor $\ell$ at the source
\begin{equation}
\xi_{\bar \nu_\ell} \equiv \frac{\phi_{\bar \ell}}{\phi_e+\phi_\mu+\phi_{\bar e}+\phi_{\bar \mu}} \, \,
\label{equ:flvini}
\end{equation}
taking into account neutrinos and antineutrinos separately, and assuming that $\phi_\tau=\phi_{\bar \tau}=0$.
The corresponding quantity $\xi^f_{\bar \nu_\ell}$ at the detector includes flavor mixing, \ie, 
\begin{equation}
\xi_{\bar \nu_\ell}^f \equiv \frac{\phi^f_{\bar \ell}}{\phi^f_e+\phi^f_\mu+\phi^f_\tau+\phi^f_{\bar e}+\phi^f_{\bar \mu}+\phi^f_{\bar \tau}} = \frac{\phi^f_{\bar \ell}}{\phi^f_{\mbox{\tiny 3f}}}\, ,
\label{equ:flvafter}
\end{equation}
where $\phi^f_{\mbox{\tiny 3f}}$  denotes the all-flavor flux. Similar quantities can be defined for any flavor and polarity.

For the charged pion decay chains, we immediately find from \equ{PPT2by2}
\begin{eqnarray}
\label{eq:pi-plus}
\pi^+
& \ra & e^+ \nue\numu\numubar \stackrel{\rm mix}{\lra} \xi_{\nuebar}^f =\frac{1}{3}\times\frac{2}{9}= \frac{2}{27} \, , \\
\label{eq:pi-minus}
\pi^- 
& \ra & e^- \nuebar\numu\numubar  \stackrel{\rm mix}{\lra}  \xi_{\nuebar}^f= \frac{1}{3} \times \left( \frac{5}{9}+\frac{2}{9} \right) = \frac{7}{27} \, .
\end{eqnarray}
We observe from the ratio of the two processes that the $\pi^-$ decay chain yields $7/2$ times more Earthly $\nuebar$ than the $\pi^+$ decay chain. From a different perspective, the Glashow event rate from the $\pi^+$ decay chain is potentially contaminated by $\pi^-$ production (if present at the source), namely
$\sim 7/2$ times the fraction $\pi^-/\pi^+$.

From \equ{PPT2by2} and \equ{flvafter}, we find the generalization
\begin{equation}
\xi^f_{\nuebar} \simeq \frac{5}{9} \ \xi_{\nuebar} + \frac{2}{9} \ \xi_{\numubar} \, 
\end{equation}
for the TBM scenario.
From this equation one can see the importance of muon antineutrinos to correctly evaluate the fraction of $\nuebar$ after the oscillations. 
The Earthly $\nuebar$ fraction for the popular production scenarios can be found in the fourth column of \Tab~\ref{table:nuebar}.
 To be more precise in terms of the mixing angles, one can express~\cite{Palladino:2015vna}
\begin{equation}
\xi_{\nuebar}^f = \frac{1}{3}(\xi_{\nuebar}+\xi_{\numubar})+ P_0 (2 \xi_{\nuebar} -\xi_{\numubar}) + P_1 \xi_{\numubar},
\label{equ:osc}
\end{equation}
in terms of the flavor composition at the source. 
This is an improvement on the TBM approximation used before, since it takes into account that the mixing angle $\theta_{13} \neq 0$. Moreover it permits to add easily the uncertainties related to the oscillation parameters.
The two parameters in this formula are given by
\begin{eqnarray}
\nonumber P_0& \simeq &0.109 \pm 0.005 \\ 
\nonumber P_1& \simeq &0.000 \pm 0.029 
\end{eqnarray}
for current data~\cite{nufit} and they represent an average between what is expected for normal and inverted mass ordering.
The uncertainty on $\xi_{\nuebar}^f$ can be estimated by adding the errors of the individual parameters in quadrature, \ie, 
\begin{equation}
\label{equ:nuerror}
\Delta \xi_{\nuebar}^f \simeq  \sqrt{(\Delta P_0 (2 \xi_{\nuebar} -\xi_{\numubar}))^2 + (\Delta P_1 \xi_{\numubar})^2} \, .
\end{equation}
 This is a simplified procedure that permits to include the uncertainties due to oscillations in an analytical manner and it should be sufficiently accurate for our purpose.

\subsection{Glashow Event Rate}
\label{sec:glashow}

The signal obtained from the $W^-$ decay can either be a cascade with deposited energy around 6.3 PeV (hadronic), a cascade with a lower deposited energy (leptonic), or a muon track with a lower energy (leptonic). 
Since a leptonic event necessarily includes an energy loss to an escaping neutrino, it cannot be distinguished from a non-resonant event at a lower energy.  Consequently, we focus on the hadronic scenario.
The resonant cross section for $\nuebar+e^-\rarr W^-\rarr {\rm hadrons}$ is 
\begin{equation}
\label{eq:Breit-Wigner}
\begin{split}
\sigma_{\rm Res}(s) = 24\pi\,\Gamma_W^2\;{\rm B}(W^-\rarr\nuebar e^-)\,{\rm B}(W^-\rarr{\rm had})   \times \\
\frac{(s/M_W^2)}{(s-M_W^2)^2 +(M_W \Gamma_W)^2}\,,
\end{split}
\end{equation}
where $M_W$ is the $W$ mass (80.4~GeV), $\Gamma_W$ is the $W$'s FWHM (2.1~GeV),
and ${\rm B}(W^-\rarr\nuebar e^-)$ and ${\rm B}(W^-\rarr{\rm had)}$ are $W^-$ branching ratios
into the $\nuebar e^-$ (10.6\%) [roughly equal to the naive 1/(3 lepton channels + 3 colors $\times$ 2 open quark channels)] and the hadronic states (67.4\%) [ $(3\times2)/(3 + 3 \times 2)$], respectively.  
At the peak, one has
\begin{eqnarray}
\sigma_{\rm Res}^{\rm peak} & = & \frac{ 24\pi\,{\rm B}(W^-\rarr\nuebar e^-)\,{\rm B}(W^-\rarr{\rm had}) }{M_W^2}  \nonumber \\
& = & 3.4\times 10^{-31} \, {\rm cm}^2\,.
\label{eq:BWpeak}
\end{eqnarray}
Consequently, the resonant cross section may be written as 
\begin{equation}
\label{eq:BW2}
\sigma_{\rm Res} (s) = \left[ \frac{\Gamma_W^2\,s}{(s-M_W^2)^2 +(M_W \Gamma_W)^2} \right]\,\sigma_{\rm Res}^{\rm peak}\,. 
\end{equation}
The $W$ width is small compared to the $W$ mass ($\Gamma_W/M_W=2.6\%$), 
and the experimental resolution will always exceed the $W$ width by far.
Thus, we can use the ``narrow width approximation'' (NWA), which is simply 
\begin{equation}
\label{eq:NWA}
\sigma_{\rm Res}(s) = \pi \Gamma_W s/M_W\,\sigma_{\rm Res}^{\rm peak}\,\delta(s-M_W^2)\,. 
\end{equation}
\begin{equation}
\label{eq:nuebar-mfp}
\lambda_{\nuebar} \sim\frac{1}{n_e\,\sigma_{\rm Res}^{\rm peak}} \sim \left\{ 
\ba{ll}
 110  \, {\rm  km} & \text{in mantle rock} \, , \\
      &  	    				\\
 310 \, {\rm  km} & \text{in ice} \, .
\ea
\right.
\end{equation}
The width in $E_\nu$, and therefore the bulk of the absorption, extends from 6.3~PeV to $\pm (2\Gamma_W)/M_W \,E_\nu$, the latter equals to $\pm 0.3$~PeV.
This short mfp, traceable to the large resonance cross section, tells us that the $\nuebar$ absorption by Earth matter at the Glashow energy of $6.3~$PeV is considerable.  
Using the Sagitta relationship between the depth $z$ of IceCube and the length of the horizontal burden $h$,
$h=\sqrt{2R_\oplus z}$, one finds an $h$ of 113-160~km for the IceCube depth 1-2~km,
well matched to the $\nuebar$ mfp.
The absence of significant overburden, the relatively short mfp of Glashow $\nuebar$'s, and the large solid angle imply that the Glashow events come mainly from horizontal directions.

The number of Glashow events can be calculated as
\begin{equation}
N^{\mbox{\tiny G}}=4\pi \times T \times \int_0^{\infty} \xi^f_{\nuebar} \phi^f_{\mbox{\tiny 3f}} \ A_{\mbox{\tiny eff}}^{\mbox{\tiny G}}(E_\nu) \ \rmd E_\nu,
\end{equation}
where the all-flavor flux $\phi^f_{\mbox{\tiny 3f}}$ is in [${\rm GeV^{-1} \, cm^{-2} \, s^{-1} \, sr^{-1}}$], $T$ is the observation time, and $A_{\rm eff}^G$ is the neutrino effective area for the Glashow events.
IceCube has published effective areas for $e$, $\mu$ and $\tau$ flavors, averaged over the neutrino and antineutrino contributions. From these one may obtain the effective area for resonance production of $\nuebar$ by 
$A_{\rm eff}^G= 2 \ (A_{\rm eff}^e-A_{\rm eff}^\tau)$, since the $\nue$ and $\nutau$ have similar effective areas except for the $\nuebar$ resonant contribution; the full explanation is reported in Appendix C. Note that the effective area includes the absorption of upgoing events;
for details, see also \App~\ref{app:crosscheck}. The Glashow event rate can be interpreted as Poissonian distributed model indicator, where we assume that the energy of the hadronic cascade produced by the decay of the $W^-$ boson is entirely detectable, \ie, the energy reconstruction is 100\% efficient and basically no contamination from ordinary deep inelastic scattering processes is expected.

The three-flavor neutrino flux $\phi^f_{\mbox{\tiny 3f}}$ at the detector is usually given in the form
\begin{equation}
\phi^f_{\mbox{\tiny 3f}} =\phi_0 \times 10^{-18} \frac{1}{\rm GeV \ cm^2 \ s \ sr} \left(\frac{E}{100 \ \rm TeV}\right)^{-\alpha}
\end{equation}
in terms of two parameters: $\phi_0$ is the normalization of the flux at $E_0=100$~TeV and $\alpha>0$ is the differential spectral index. We will give the values of $\phi_0$ of the different datasets in the following. 

Since the flux of neutrinos is fitted by a power law and the Glashow resonance cross section is approximately a Dirac $\delta$-function, it is possible to derive a useful analytical expression for the expected number of events $N^{\mbox{\tiny G}}$ as
\begin{eqnarray}
N^{\mbox{\tiny G}} & \propto & \int_0^\infty \left(\frac{E}{E_0}\right)^{-\alpha}  \delta{(E-E^*)} \ \rmd E \nonumber \\
& = & \left( \frac{E^*}{E_0}\right)^{-\alpha} = e^{-\alpha \log(E^*/E_0)} \, , \nonumber
\end{eqnarray}
where $E^*=6.32$ PeV. We find numerically, around the Glashow resonance,
\begin{equation}
N^{\mbox{\tiny G}}  \simeq  \xi_{\nuebar} \times \phi_0  \times T_{\mbox{\tiny exp}}  \times 1.071 \times \exp \left(-\frac{\alpha-2}{0.244} \right) \, ,
\label{equ:rateG}
\end{equation} 
where $T_{\mbox{\tiny exp}}$ is the exposure in units of years using the present effective area of the completed IceCube detector provided in \Ref~\cite{aeffweb}.
This means, one year of IC86 operation corresponds to $T_{\mbox{\tiny exp}}=1$, and at present, the exposure can be estimated to be $T_{\mbox{\tiny exp}} \simeq 4$. We refer to $T_{\mbox{\tiny exp}}$ as {\bf ``IC86 equivalent exposure''} in the following.

\subsection{Methods}

Using the ``IC86 equivalent exposure'' $T_{\mbox{\tiny exp}}$ in \equ{rateG}, we evaluate the expected number of events and their uncertainties (assumed to be statistics-limited). This  procedure  allows to extrapolate our results easily to the next-generation detectors, such as IceCube-Gen2. Indeed this detector should have an exposure from 5 to 12 times greater than IC86 (see \fig~10 of \Ref~\cite{Aartsen:2014njl}), which means that about 50-120 years of IC86 equivalent exposure can be accumulated after ten years of Gen2 operation. Note that the precise exposure will depend on the currently ongoing detector optimization; we therefore show this range in our figures. Moreover, note that the required exposures for different spectral indices can be found by using \equ{rateG}, but the normalization $\phi_0$ has to be known; see \eg\ \fig~1 of \Ref~\cite{Aartsen:2015knd} for the correlation between the spectral index and the normalization.

The Through-going muon analysis indicates a spectral index $\alpha \simeq 2$~\cite{Aartsen:2015zva,Aartsen:2016xlq}.  Under the hypothesis of the pion decay chain, which approximately gives flavor equipartition after flavor mixing, one obtains the all-flavor flux by multiplying the muon neutrino flux by a factor three. Since the spectral index slightly changes at each analysis we assume $\alpha=2$, that is also expected from simplest shock acceleration of the primaries. From \fig~6 of \Ref~\cite{Aartsen:2016xlq}  it is possible to estimate the all flavor normalization at 100 TeV of the $\phi_{\numu}+\phi_{\numubar}$ flux for $\alpha=2$, which is   $\phi_0 \simeq 2$. In this case the expected number of events are about 0.35 and 0.17 per year for the reference pp and ${\rm p}\gamma$ scenarios, respectively. Different analyses of the astrophysical neutrinos, however, yield different results.

Using the best-fit of the high-energy starting events (HESE)~\cite{Aartsen:2015knd}, one finds  $\phi_0=6.7$ and $\alpha=2.5$. Here the expected number of Glashow resonance events is reduced by a factor $\simeq$ 2.5, which can be obtained from \equ{rateG}. 

In the figures of this study we always use the hard flux obtained from the Through-going muon events, with a spectral index $\alpha \simeq 2$. 
Note that the hard spectrum, suggested by Through-going muons, is in contradiction with the softer spectrum suggested by the global analysis of HESE.  It is plausible that at low energy a Galactic component, mainly seen from the Southern hemisphere due to the position of the Earth in the Galaxy, can increase the spectral index up to the observed $E^{-2.5}$, whereas at high energy, the extragalactic component is predominant and the spectrum becomes harder $\sim E^{-2}$~\cite{Palladino:2016zoe,PV2}.
Moreover, the $E^{-2}$ spectrum is perfectly compatible with the current observation of three events above 1 PeV, and a spectral hardening at higher energies is currently investigated by the IceCube collaboration~\cite{KowalskiNeutrino2016}. We will in some cases also refer to the HESE best-fit flux for completeness.
Note again that the required IC86 equivalent exposure can be easily re-scaled for different datasets using \equ{rateG}.

In order to study different scenarios, we will typically assume that the spectral index follows observations, but we take the flavor and neutrino-antineutrino compositions at the source from specific model predictions. From the model perspective, this procedure is strictly speaking inconsistent, as one would like to describe (fit) spectrum and flavor composition at the same time. On the other hand, the pion spectrum in photohadronic models depends on the shape of the target photon spectrum. Therefore, in general one can not expect a simple power law that can be easily compared to existing experimental results. We nevertheless selected examples where the spectral shape matches the characteristics of available data, such that they can be compared with certain source ``prototypes''. For example, sources with strong magnetic fields will suffer from muon damping at the highest energies, and sources with high photon densities from optical thickness to photohadronic interactions. 

Since viable models exist for Gamma-Ray Bursts (GRBs) which can predict the flavor composition and neutrino-antineutrino composition, we have chosen these in some cases. Although long-duration GRBs are ruled out as dominant contribution to the observed diffuse flux~\cite{Abbasi:2012zw}, low-luminosity or hidden GRBs are possible candidates~\cite{Murase:2008mr,Senno:2015tsn,Tamborra:2015qza}. In addition, we expect similar characteristics for other event classes, such as tidal disruption events~\cite{Wang:2011ip,Lunardini:2016xwi}.

\section{Results for standard scenarios}
\label{sec:exp}

This section contains the results for the ideal pp and ${\rm p}\gamma$ scenarios, which include realistic decay kinematics of the pion and muon decays and the mixing angles for flavor mixing. We also compare with results from Monte Carlo simulations of charged pion production. 

\subsection{Ideal scenarios for charged pion production}

In order to define our {\em ideal} pp and {\em ideal} ${\rm p}\gamma$ scenarios we continue to use pion production according to \equ{delta} and \equ{idealpp}.  However, in addition to the reference models, we take into account the energy distributions from pion and muon decays assuming a power law pion spectrum, and we include realistic flavor mixing using the mixing angles in \Ref~\cite{nufit}. Note that the power law index of pions follows that of the protons in the pp case, while it depends on the target photon spectrum for ${\rm p}\gamma$ interactions in general. For power law spectra, we can quantify the effect of the pion and muon decay kinematics with spectrum weighted moments, so-called $Z$ factors~\cite{Lipari:1993hd}, see \App~\ref{app:zeta}. Since the energy distribution of the three daughter neutrinos is not the same because the $\numu$ are produced in two body decays, whereas the other neutrinos are produced in three body decays, we expect small deviations from the above reference compositions. 
The results for the ideal pp and ${\rm p}\gamma$ scenarios are shown in \Tab~\ref{flvcompid} as a function of the spectral index (of the pions) $\alpha$. 
From the table we notice that especially for softer spectral indices the flavor composition of ideal ${\rm p}\gamma$ sources differs from the reference case by up to 30\% (for $\xi_{\numu}$) due to kinematics. 

\begin{table}[t]

\caption{\textit{Flavor composition at the source for our reference and ideal pp and $p\gamma$ scenarios for different spectral indices $\alpha$ of the pions. Here ``reference'' assumes that the neutrino flavors are populated according to \eqs~(\ref{equ:delta}) to~(\ref{equ:idealpp}) without taking into account the secondary decay kinematics, whereas``ideal'' refers to taking into account the pion/muon decay kinematics for power law spectra (but no deviations from the reference $\pi^+/\pi^-$ ratio at the source).}}
\label{flvcompid}
\begin{center}
\begin{tabular}{|l|c|cccccc|}
\hline
Production & $\alpha$ & $\xi_{\nue}$ & $\xi_{\numu}$ & $\xi_{\nutau}$ & $\bar{\xi}_{\nue}$ & $\bar{\xi}_{\numu}$ & $\bar{\xi}_{\nutau}$\\
\hline
Reference pp & any & 0.167 & 0.333 & 0 & 0.167 & 0.333 & 0 \\
Ideal pp & 2.0 & 0.175 & 0.325 & 0 & 0.175 & 0.325 & 0   \\
Ideal pp & 2.3 & 0.179 & 0.321 & 0 & 0.179 & 0.321 & 0 \\
Ideal pp & 2.6 & 0.183 & 0.317 & 0 & 0.183 & 0.317 & 0  \\
\hline
Reference ${\rm p}\gamma$ & any & 0.333 & 0.333 & 0 & 0 & 0.333 & 0 \\
Ideal ${\rm p}\gamma$ & 2.0 & 0.350 & 0.290 & 0 & 0 & 0.360 & 0 \\
Ideal ${\rm p}\gamma$ & 2.3 & 0.358 & 0.273 & 0 & 0 & 0.369 & 0  \\
Ideal ${\rm p}\gamma$ & 2.6 & 0.366 & 0.256 & 0 & 0 & 0.378 & 0   \\
\hline
\end{tabular}
\end{center}
\end{table} 

\begin{figure*}[t]
\centering
\includegraphics[width=0.4\linewidth]{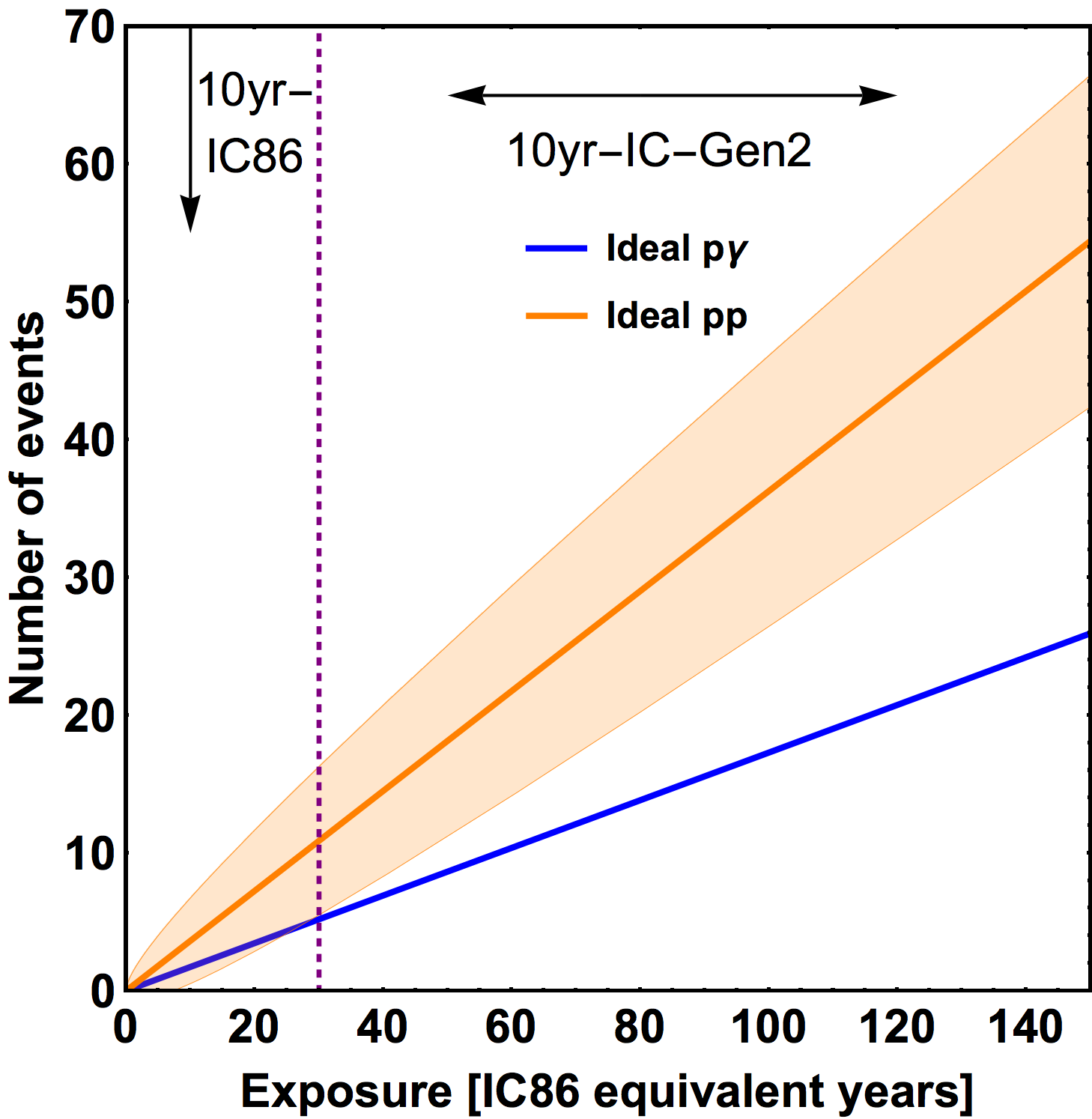} \hspace{1.0cm}
\includegraphics[width=0.4\linewidth]{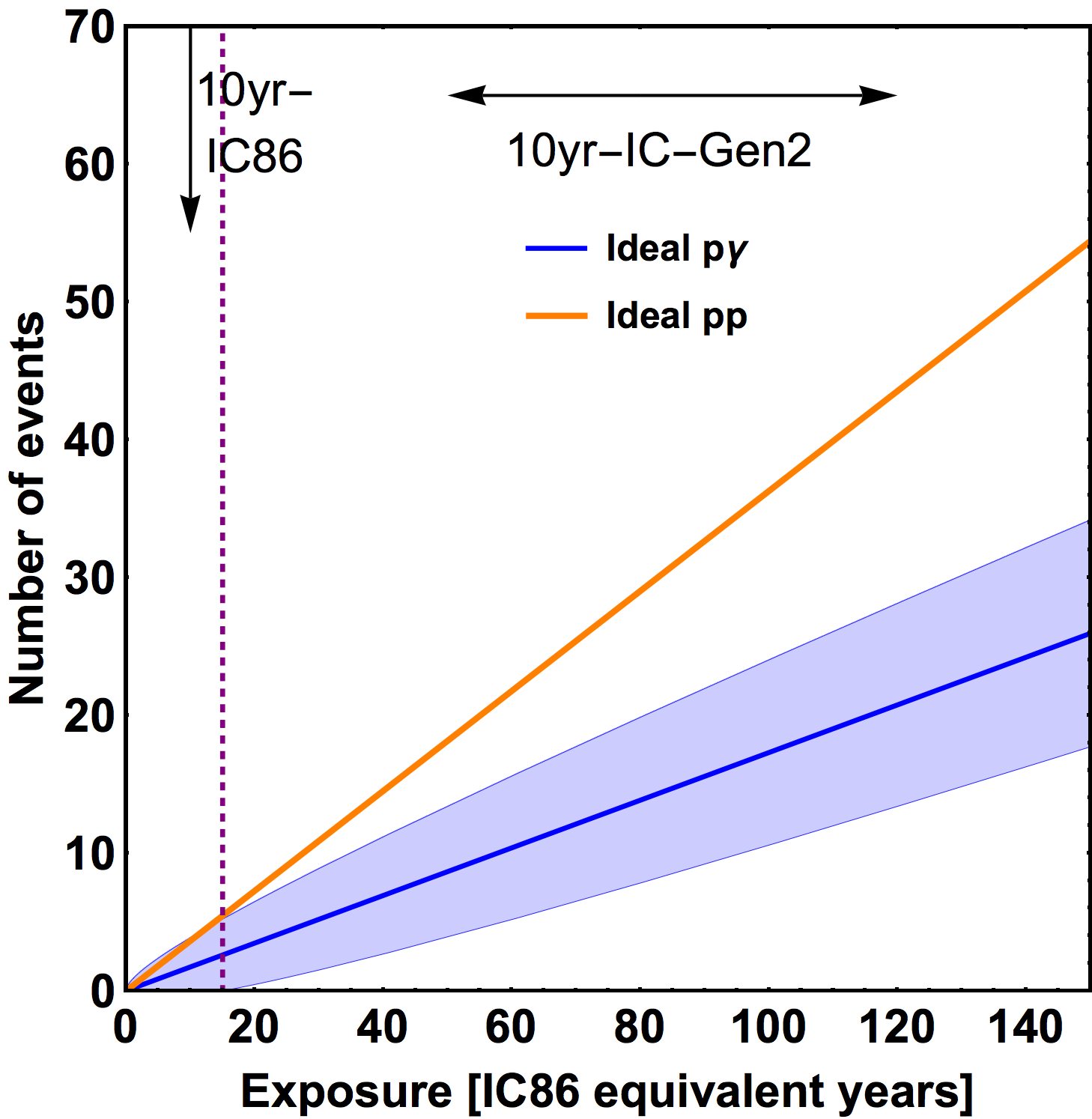}
\caption{\textit{Expected number of Glashow events in the ideal pp and $p\gamma$ scenarios as a function of the exposure for $\alpha=2.0$. The bands represents the 90\% C.L. interval from the statistical (Poissonian) uncertainty and the model uncertainties on the oscillation parameters, assuming a true pp and $p\gamma$ scenario in the left and right panel, respectively. The vertical lines indicate when the other scenario can be excluded.}}
\label{fig:ideal}
\end{figure*}

Let us now introduce our concept of exposure using very simple examples, and consider the ideal pp and ${\rm p}\gamma$ cases only. We show in \figu{ideal} the expected number of Glashow events in the ideal scenarios as a function of the exposure, where the bands illustrate the statistical (Poissonian) error $\delta_s$ in combination with the error on the oscillation parameters $\delta_p$ at the 90\% C.L. The error from flavor mixing is small presently (about 10\%) and will become negligible with improving oscillation parameters in the next generation of detectors. We assume that their relative error scales $\propto$~$1/\sqrt{T}$ with exposure. The total error $\delta$ is obtained as $\delta=\sqrt{\delta_s^2+\delta_p^2}$.

In the left panel of \figu{ideal}, it is assumed that the true source (data) is an ideal pp source. In that case, the shaded region determines the error, and that region separates from the solid ${\rm p}\gamma$ curve for $T_{\mbox{\tiny exp}} \gtrsim  30$ (see vertical line). This means that the ideal ${\rm p}\gamma$ source can be excluded after this exposure, which is clearly beyond IC86, but well within the reach of IceCube-Gen2 (compare to arrows). If the true source is an ideal ${\rm p} \gamma$ source, only 15 equivalent years of exposure are needed, as it is illustrated in the right panel of \figu{ideal}. We will use similar representations for less trivial examples in the following, where we typically show the shaded region for one curve only. Note, however, the exposure required to discriminate among scenarios depends on the combination of the true scenario at the source and the detected diffuse flux at Earth.

\subsection{Monte Carlo modeling of interactions}

\begin{figure}[t]
\centering
\includegraphics[width=0.9\linewidth]{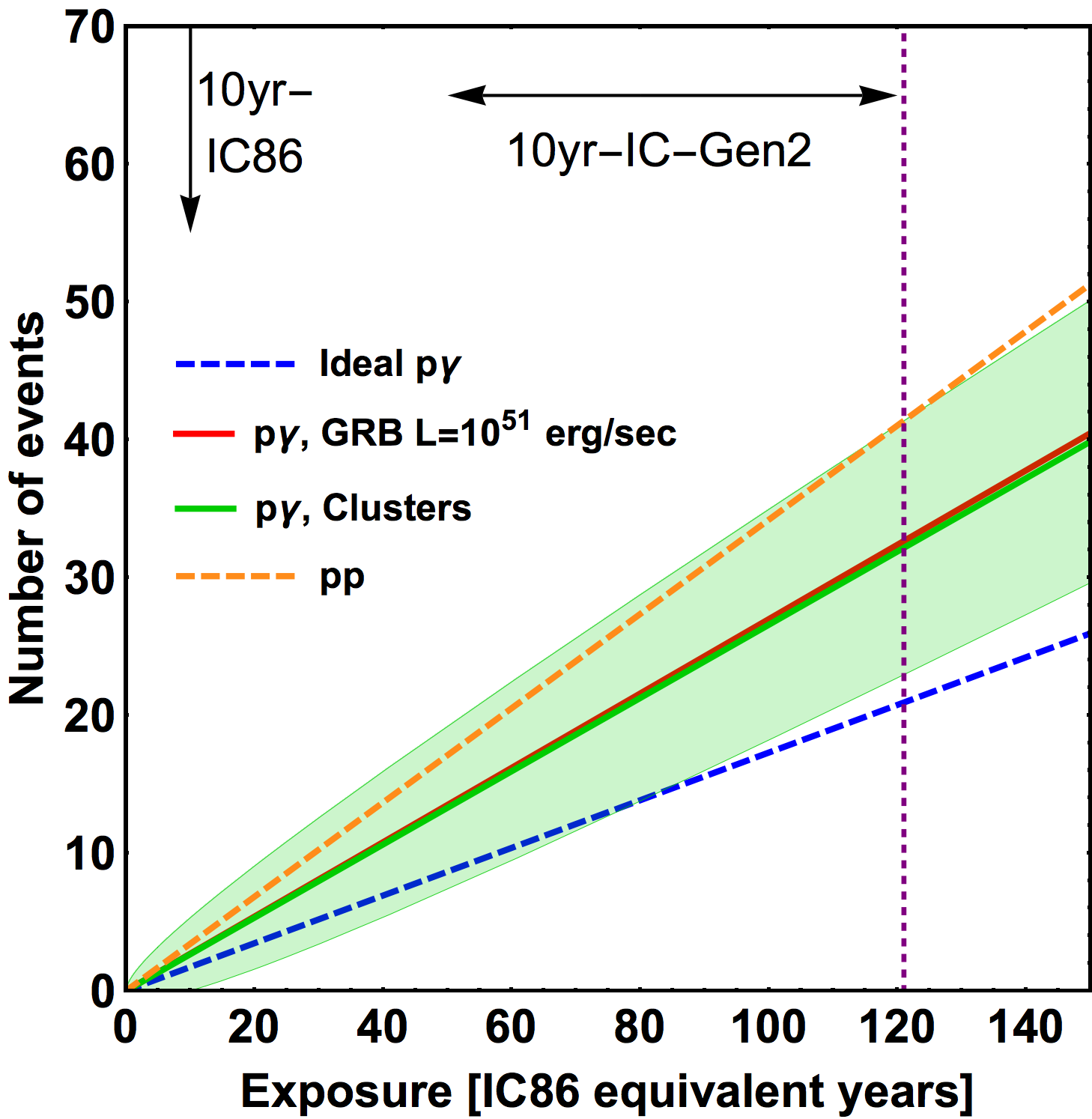}

\caption{\textit{Same as \figu{ideal} demonstrating the impact of contaminations from Monte Carlo event generators for the pion production. The different curves show the ideal $p\gamma$ case, pp and $p\gamma$ scenarios from Monte Carlo simulations of the interactions with {\sc Sibyll}, {\sc Epos-lhc} or {\sc Qgsjet} for pp (yielding similar results), and {\sc Sophia} for $p\gamma$. For the $p\gamma$ interactions, a GRB target photon spectrum with spectral indices $-1$ and $-2$ and a break at $1 \, \mathrm{keV}$ (shock rest frame) has been assumed (red curve), as well as a synchrotron target photon spectrum from co-accelerated electrons (cluster scales with $R \simeq 10^{19} \, \mathrm{km}$, $B \simeq 10^{-6} \, \mathrm{G}$, green curve).}}
\label{fig:cont1}
\end{figure}

So far, we have included the kinematics of pion and muon decays assuming power-law pion spectra, but we have used the reference values for the $\pi^-/\pi^+$ ratio for pp (ratio $1$) and ${\rm p}\gamma$ (ratio $0$) interactions from \eqs~(\ref{equ:delta}) and~(\ref{equ:idealpp}). As we demonstrate, this result is not what we obtain from more realistic Monte Carlo simulations of the interactions.

For pp interactions in astrophysical densities, the spectral index of the secondary pions is equal to the projectile spectrum, since there is no absorption by the surrounding medium and the target particles are non-relativistic. The usual assumption of isospin symmetry, where $\pi^+$, $\pi^-$ and $\pi^0$ are produced in equal quantities, does not hold in detailed Monte Carlo simulations when using high-energy hadronic interaction models {\sc Sibyll} 2.3 \cite{Ahn:2009wx,Riehn:2015tbd}, {\sc Epos-lhc} \cite{Pierog:2013ria} and {\sc Qgsjet}-II-04 \cite{Ostapchenko:2010vb}. These predictive Monte Carlo models are widely used in simulations of cosmic ray interactions, minimum-bias and forward physics studies at the LHC \cite{Akiba:2016ofq}. Due to the power-law spectrum of the interacting nucleons, the secondary particles carrying a large fraction of the projectile's energy $x \gtrsim 0.3$ (in the projectile fragmentation zone), are important for the computation.\footnote{This can be seen by comparison with the integrand of \equ{zeta_factor}.} At typical energies of hundreds of PeV in our present kinematics, these large momentum fractions of secondaries mostly involve valence quark scatterings and their subsequent hadronization. Given $u$ valence quark dominance in protons, one naturally expects $\pi^+/\pi^- > 1$ in the forward phase-space. Depending on the spectral index, that selects the relevant part of the particle production phase-space, the $Z$ factors for pp collisions can substantially deviate from the prediction $\pi^-/\pi^+ \simeq 1$, whereas the kinematics of the pion and muon decays alleviates the problem somewhat. At the end, the difference compared to the reference pp case is up to  16\% ($\alpha=2$), 22\% ($\alpha=2.3$), and 30\% ($\alpha=2.6$). We discuss this issue in detail in \App~\ref{app:zeta}. Note that in the following, we will only show the pp curve including the Monte Carlo simulation, where the result hardly depends on which of the three  Monte Carlo event rate generator is used. For a discussion of the effect of kaons, which is however small, see \App~\ref{app:kmesons}. 

Photohadronic interactions do not only include the $\Delta$-resonance in \equ{delta}, but also direct ($t$-channel) pion production, higher resonances, and high-energy multi-pion processes~\cite{Mucke:1999yb} -- especially the latter lead to almost $\pi^-/\pi^+ \simeq 1$; see \eg\ \Refs~\cite{Hummer:2010vx,Baerwald:2010fk} for an illustration of the individual contributions. Since these processes lead to $\pi^-$ production as well, an intrinsic contamination with $\pi^-$ is expected. This contamination depends on the target photon spectrum, and can lead to a $\nuebar/\nue$ ratio at the source of about 25\%-50\% (AGNs/GRBs) to 100\% (10~eV thermal target photon spectrum); see Fig.~10 in \Ref~\cite{Hummer:2010vx}.\footnote{Note that the $\nuebar$ alone are not sufficient to describe the impact on the Glashow resonance, as $\numubar$ may mix into $\nuebar$ - as we discussed earlier.} As an additional complication, since the pion spectrum depends on the photon spectrum and is, in general, not a simple power law, the pion and muon decays have in this case been computed numerically taking into account the re-distribution functions of the secondaries~\cite{Lipari:2007su}. 

Here we pick two representative examples in the middle of extremes. In \Ref~\cite{Hummer:2010ai}, the target photons are assumed to be generated by the synchrotron radiation of co-accelerated electrons, which is a typical assumption for AGNs. This model has been fit to IceCube data in \Ref~\cite{Winter:2013cla}, where it has been demonstrated that the indication for a cutoff at PeV energies can be interpreted in terms of a limited maximal proton energy or strong magnetic field effects. We pick one benchmark point (TP8, size of acceleration region $R \simeq 10^{19} \, \mathrm{km}$, $B \simeq 10^{-6} \, \mathrm{G}$) from that scenario with a sufficiently large proton energy to allow for Glashow resonant events, which at the same time implies that photohadronic contaminations cannot be avoided because of the high available center-of-mass energies. The parameters for this benchmark correspond to the scale of galaxy clusters, and the predicted flavor composition $(\xi_{\nue}: \xi_{\nu_\mu}: \xi_{\nutau} :\xi_{\nuebar}: \xi_{\bar \nu_\mu}: \xi_{\bar  \nutau})$ at the source that is equal to $(0.27: 0.32 : 0: 0.09: 0.32 :0)$. As an alternative, we present a GRB example~\cite{Baerwald:2011ee,Hummer:2011ms} representative for sources with stronger magnetic fields. Here the parameters have been chosen {\em not} to be in the muon damped regime at the Glashow resonance\footnote{The parameters are  $L_\gamma=10^{51} \, \mathrm{erg \, s^{-1}}$, variability timescale $t_v=0.1$~s, Lorentz factor $\Gamma = 300$, redshift $z = 2$,  and a broken power law photon spectrum with photon break $\varepsilon_b = 1$ keV in the shock rest frame, first spectral index $\alpha = -1$ and second spectral index $\beta = -2$. A smaller value for $t_v$ would lead to spectral cooling breaks dominated by adiabatic losses, and to muon cooling at the Glashow resonance; see \fig~\ref{mudamp}.}  -- a case which we discuss below; the flavor composition at the source is $(0.27: 0.31 : 0: 0.09: 0.33 :0)$.

In \figu{cont1} we show a comparison among the Monte Carlo results and the ideal ${\rm p} \gamma$ case. The two ${\rm p} \gamma$ examples are, in spite of very different astrophysical environments, very similar in this case. The discrimination from the pp sources requires about 120 equivalent years of exposure, which is at the upper end of the expected 10yr exposure of IceCube-Gen2, and therefore challenging. In fact, one sees that the ideal ${\rm p}\gamma$ case can be excluded already after 81 equivalent years. Note, however, that the ${\rm p} \gamma$ prediction depends on the model parameters, and even within the model in \Ref~\cite{Hummer:2010ai} a wide range of possibilities is predicted. The contamination from multi-pion processes seems to depend on the maximal proton energy, as higher proton energies allow for higher center-of-mass energies, where multi-pion processes dominate. The fact that we need to have neutrinos at 6.3~PeV implies that the proton energy has to be at least around 120~PeV. We observe  that sources which produce neutrinos with Glashow resonance energies tend to have large multi-pion contributions, and the chosen contaminations are representative. We will henceforth use the GRB case as prototype for the ${\rm p}\gamma$ source. 

\section{Impact of source conditions}
\label{sec:cond}

In this section we demonstrate how the source conditions can affect the Glashow event rate, and whether different scenarios can be discriminated based on the Glashow resonance only. Examples are optically thick (to photohadronic interactions) sources, sources of heavier nuclei as primaries for the neutrino production, and sources with strong magnetic fields leading to muon damping at the Glashow resonance.

\subsection{Optically thick sources}

\begin{figure*}[t]
\centering
\includegraphics[width=0.4\linewidth]{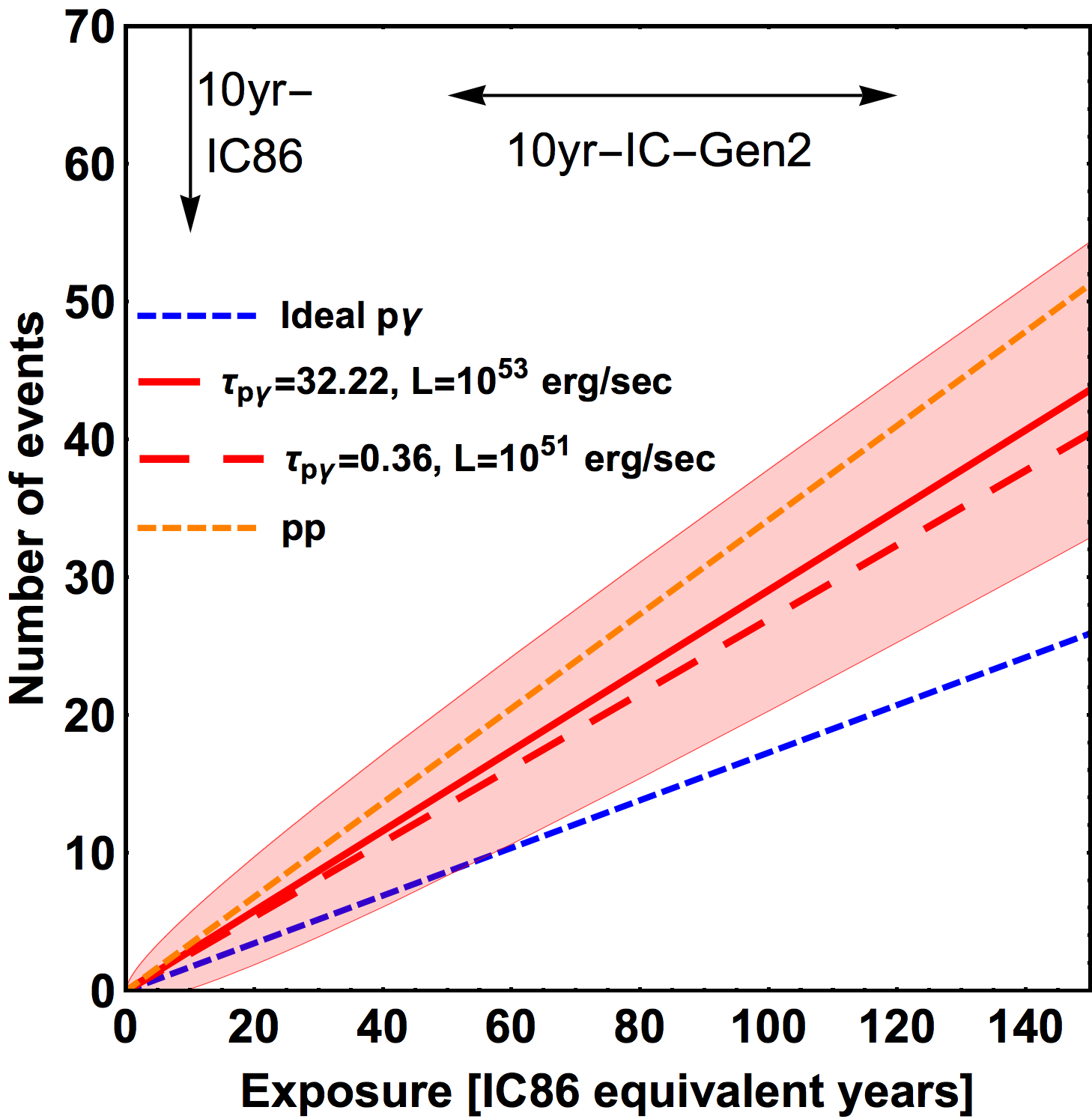} \hspace{1.0cm}
\includegraphics[width=0.45\linewidth]{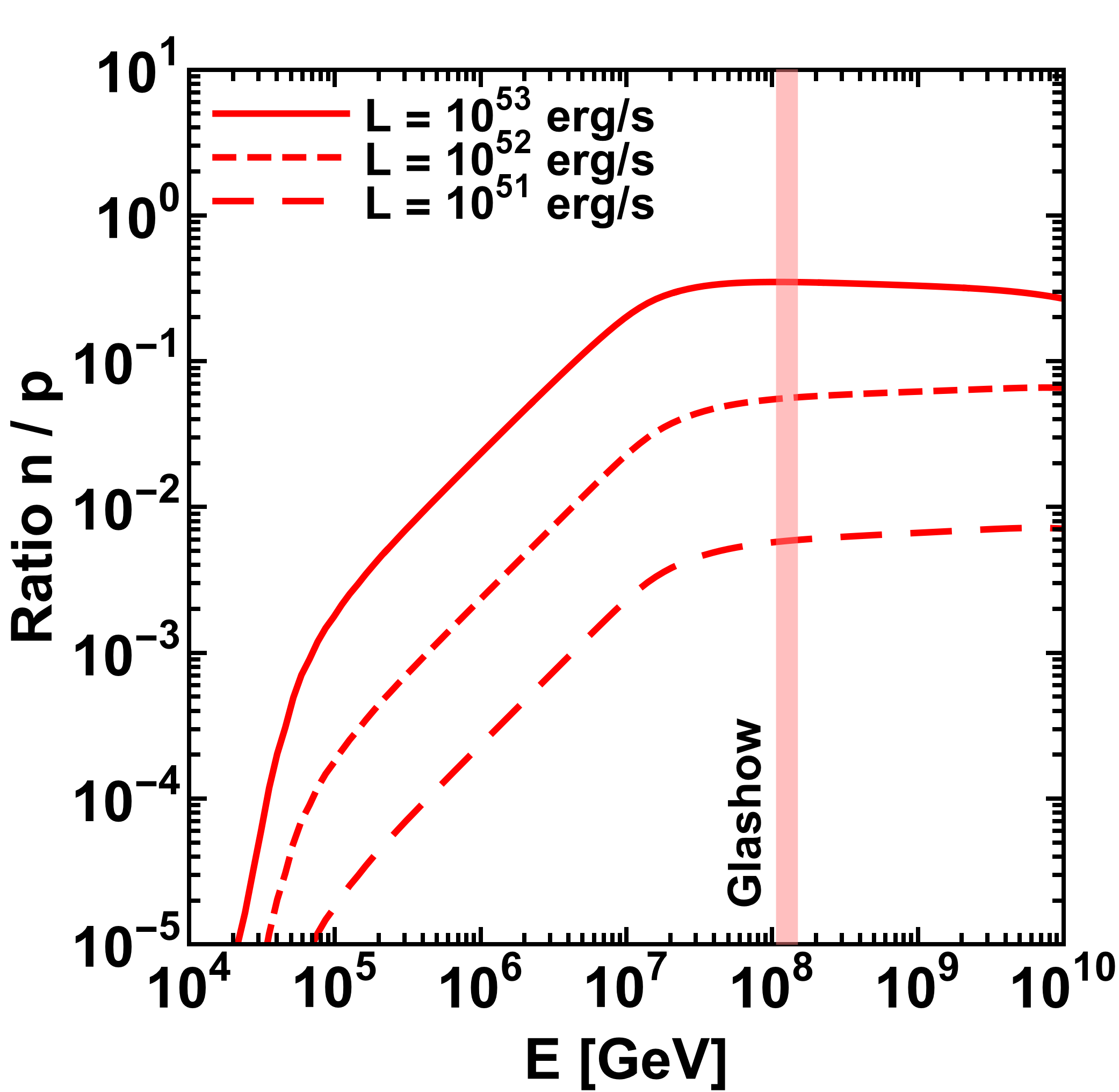}

\caption{\textit{Left panel: expected number of Glashow events as a function of exposure for the GRB case for varying optical thickness to photohadronic interactions $\tau_{p \gamma}$. As the luminosity in the burst increases, the optical thickness increases as well, leading to an increasing contamination by $\pi^-$.  Right panel: neutron to proton ratio as a function of the energy for different luminosities. At the Glashow energy (the vertical band indicates the corresponding primary energy), the ratio scales linearily with the luminosity, saturating at approximately 30\% for $L = 10^{53}$ erg/s. }} 
\label{ponly}
\end{figure*}

 From \equ{delta}, it is obvious that photohadronic interactions will produce neutrons. If the source is  optically 
thick to photohadronic interactions, \ie,  $\tau_{p \gamma} \equiv d'/\lambda'_{\text{mfp}} \gg 1$, where $d'$ is the shell thickness (source size) and $\lambda'_{\text{mfp}}$ the mean free path 
(shock rest frame), the nucleons will interact multiple times before leaving the source. 
The leading $\Delta$-resonance for neutrons is isospin-symmetric to protons  
 \begin{equation}
n+\gamma \rightarrow \Delta^0 \rightarrow \begin{cases} \pi^- + p &\text{ 1/3 of all cases} \\ \pi^0 + n &\text{ 2/3 of all cases}\end{cases} \, ,
\label{equ:deltan}
\end{equation}
which means that $\pi^-$ are pre-dominantly produced instead of $\pi^+$. In the optically thick case, one roughly expects a neutron energy spectrum which is about $1/3$ (branching ratio) times $0.8$ (fraction of primary energy deposited into the secondary nucleon) $\simeq 30\%$ that of the inital proton spectrum, with a corresponding fraction of $\pi^-$ contamination at the highest energy.

We use the GRB example from the previous subsection to demonstrate the effects of optical thickness, where $\tau_{p \gamma}$ is evaluated at the maximal proton energy; see  \Ref~\cite{Baerwald:2013pu} for a discussion on the impact of the optical thickness.  However, we update this simulation with an explicit treatment of the coupled proton-neutron system, \ie, we solve the coupled time-dependent partial differential equation system; details are presented in \Refs~\cite{Boncioli:2016lkt,grbprep}.
An increasing optical thickness in this model is obtained by increasing $L_\gamma$.
Note that while the observed GRBs may not be powering the diffuse neutrino flux, similar conclusions will apply to AGNs and other optically thick sources.
We present the neutron to proton ratio as a function of (observed) energy in Fig.~\ref{ponly}, right panel, for different luminosities. In the saturation case (highest energy), we find about the expected 30\% neutron contamination. 

In the left panel of Fig. \ref{ponly}, we show that the expected Glashow event number, \ie, the fraction of $\nuebar$ after mixing, increases with optical thickness. This is supported by the right panel of Fig. \ref{ponly}. Since this causes a significant contamination by $\pi^-$, 
 the difference between the pp and ${\rm p}\gamma$ mechanisms is reduced, making it harder to distinguish among them. Not even an equivalent exposure of 150 years is sufficient to distinguish between a pp source and a ${\rm p}\gamma$ source in which a strong optical thickness is present. The fraction of $\nuebar$ is about 10\% more in the case of large optical thickness with respect to the standard ${\rm p}\gamma$ case. This is sufficient to reduce the difference between ${\rm p}\gamma$ and pp mechanism and to significantly increase the required exposure to distinguish among them; this makes the discrimination from a pp source impossible, even with a next generation detector. 

\subsection{Heavy nuclei primaries} 

\begin{figure*}[t]
\centering
\includegraphics[width=0.4\linewidth]{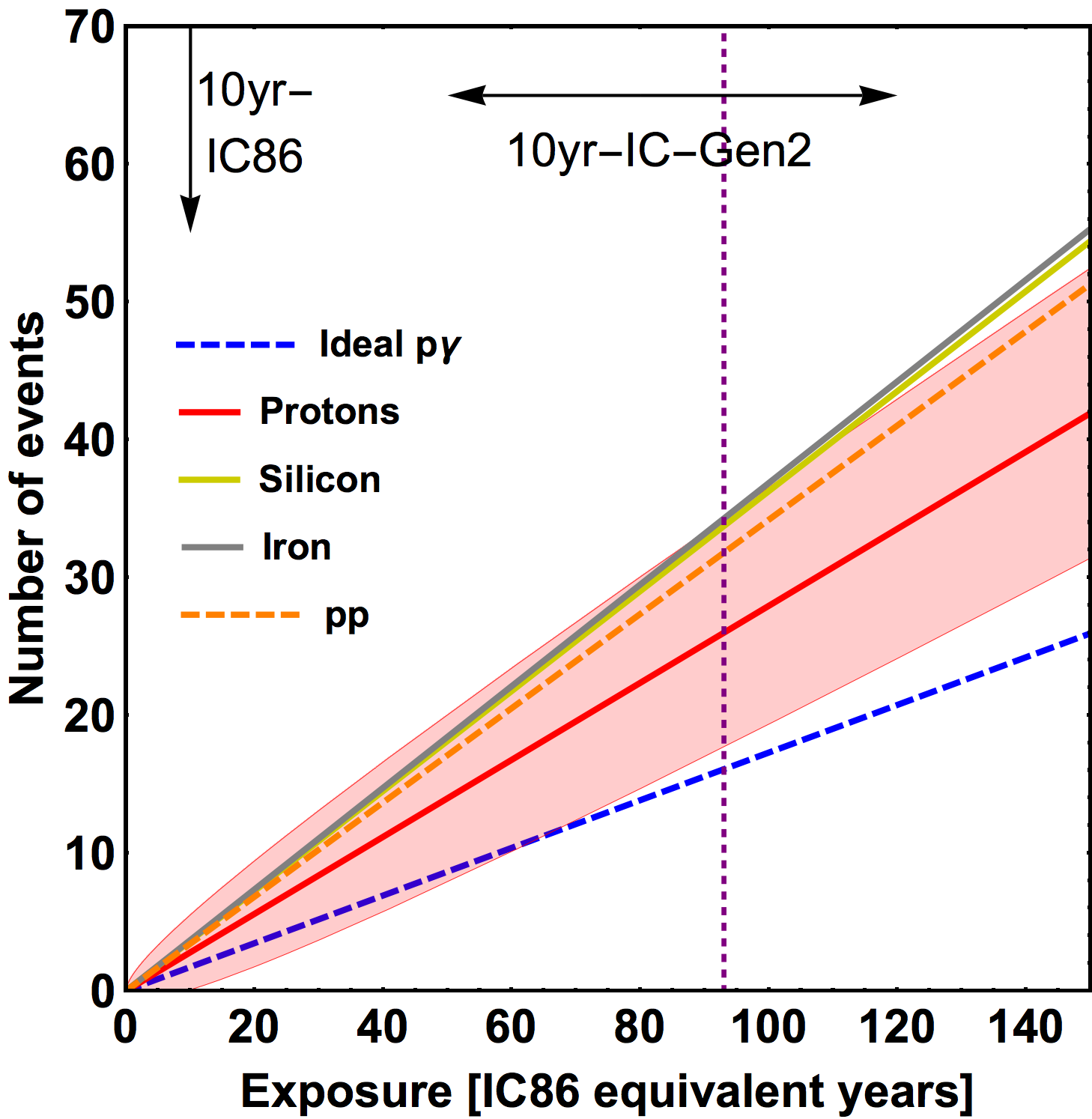} \hspace{1.0cm}
\includegraphics[width=0.45\linewidth]{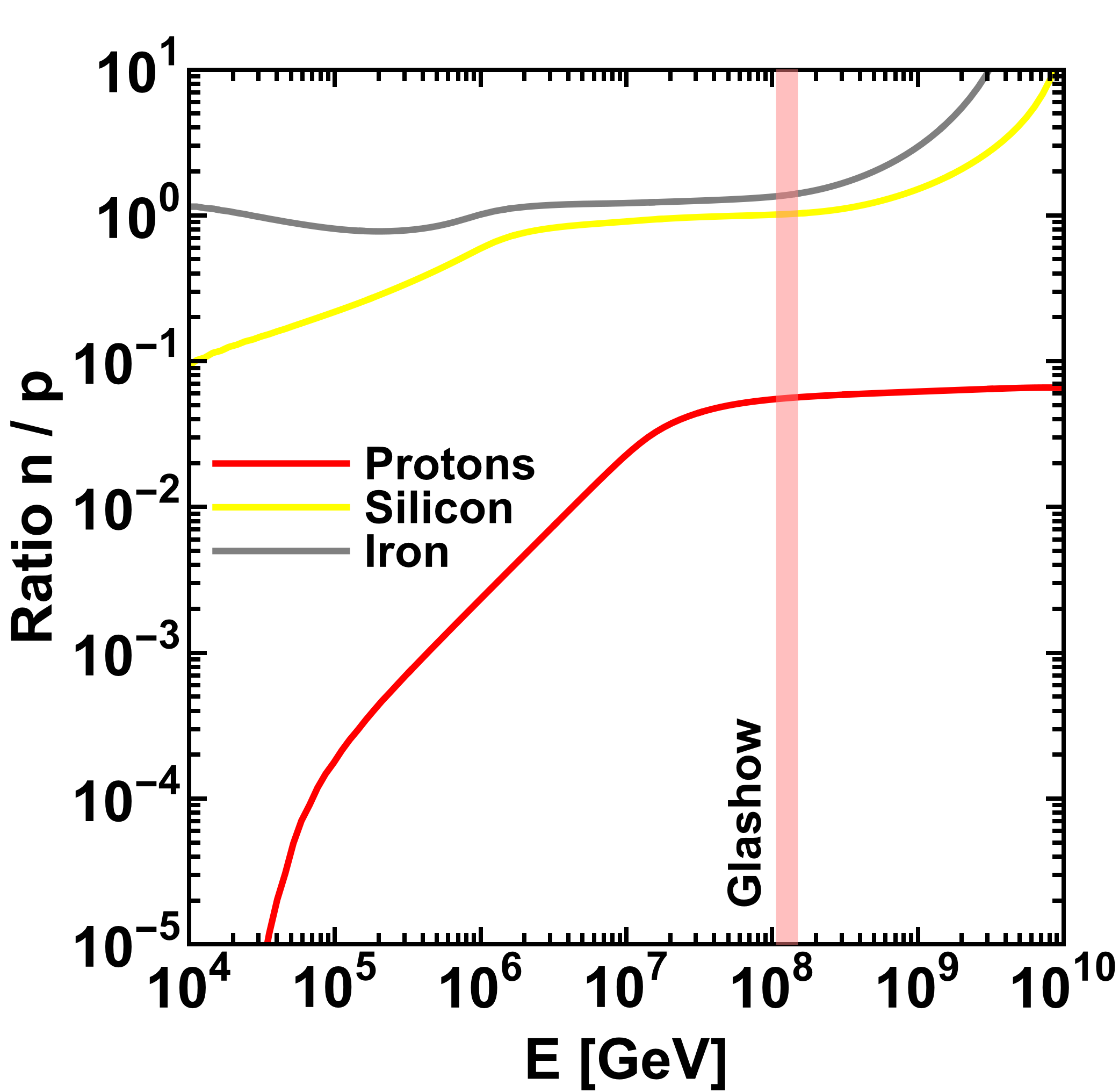}
\caption{\textit{Left panel: expected number of Glashow events as a function of exposure in the GRB case for different injected primary isotopes (pure composition assumed, photohadronic interactions). It is possible to exclude a source containing heavier isotopes from a source containing protons after about 95 equivalent years (vertical dashed line).  Right Panel: neutron to proton ratio in the source (reached in the state steady from the disintegration of nuclei)  as a function of energy for different injection isotopes. At the typical nucleon energy responsible for Glashow events ($\sim 20 \times 6.3$ PeV, indicated by vertical band) the number of interacting neutrons and protons is similar for iron and silicon. Here $L=10^{52}$ erg/s.}}
\label{fig:grb}
\end{figure*}

The nuclear composition of the primaries leading to the observed neutrino flux is highly uncertain. For example, the observed cosmic ray composition non-trivially changes as a function of energy in the relevant primary energy range, see \eg\ \Ref~\cite{Gaisser:2013bla}. Thus, neutrinos from cosmic ray interactions in our galaxy would carry the composition information as a spectral imprint~\cite{Joshi:2013aua}. Furthermore,  ultra-high energy cosmic ray observations by Auger indicate a composition significantly heavier than protons at the highest energies~\cite{Abraham:2010yv}. It is therefore possible that a part of the diffuse neutrino flux comes from interactions of heavier nuclei. 

We simulate the nuclear cascade in the GRB case following \Ref~\cite{Boncioli:2016lkt}. The implications for the Glashow resonance are shown in \figu{grb} with injection of either protons, $^{28}$Si, or $^{56}$Fe only. We again choose parameters such that muon damping does not occur at the Glashow resonance.  Note that the  proton case is slightly different compared to the standard ${\rm p}\gamma$ case, because it is obtained considering a GRB with a luminosity of $L=10^{52}$ erg/s and $t=0.1$ s.

As one important observation, nuclear disintegration of heavy nuclei leads to emission of light fragments, mostly protons and neutrons. These can contribute to the neutrino production similar to the primary nuclei~\cite{Anchordoqui:2007tn}. The ratio between neutrons and protons is approximately determined by the neutron-proton ratio of the primary nucleus, since at high energies neutrons don't decay inside the source, see right panel. Note that in the chosen example the neutrino production is dominated by the secondary protons and neutrons \cite{grbprep}, whereas for smaller luminosities the photo-meson production off heavier nuclei can dominate. We therefore show in the right panel of \figu{grb} the neutron to proton ratio produced by the disintegration of nuclei.
For iron and silicon, the expected signal is larger than in the pp scenario. This is due to the fact that these isotopes are neutron-rich (see right panel of \figu{grb}) and, as a consequence, the source contains more $\pi^-$ from the process \equ{deltan}. This is confirmed by the flavor composition at the source, that is almost the same for iron and silicon, namely
$(0.18: 0.32 : 0: 0.18: 0.32 :0)$.
Obviously, in this case there is no possibility to distinguish between pp and ${\rm A}\gamma$ even if the exposure is huge. 

It is however very interesting to observe that the $A \gamma$ scenarios can, almost independent of the composition (for $A\ge2$), be ruled out after about 95 equivalent years for a proton composition, what is potentially within the reach of IceCube-Gen2. In the inverted case of an $A \gamma$ source, the expected exposure to rule out protons is larger (124 equivalent years). If the source is known to have high radiation densities such that photohadronic interactions dominate, one can therefore use the Glashow resonance as a smoking gun signature for primary nuclei. Note that the scenario ``Protons'' in \figu{grb} is not exactly the same as our standard ${\rm p} \gamma$ scenario, as the optical thickness is slightly higher.

\subsection{Muon damped sources}

\begin{figure*}[t!]
\centering
\includegraphics[width=0.4\linewidth]{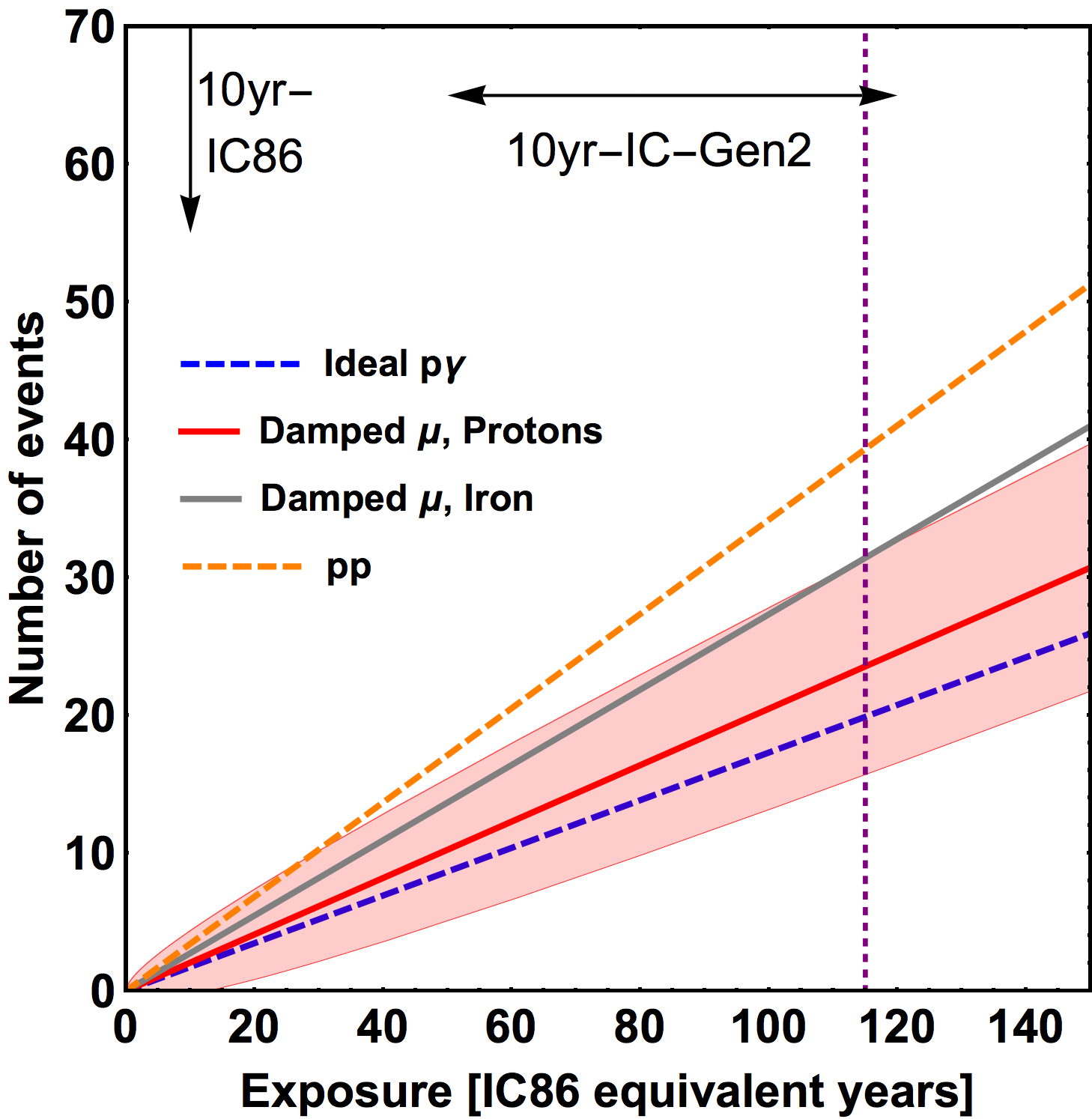} \hspace{1.0cm}
\includegraphics[width=0.425\linewidth]{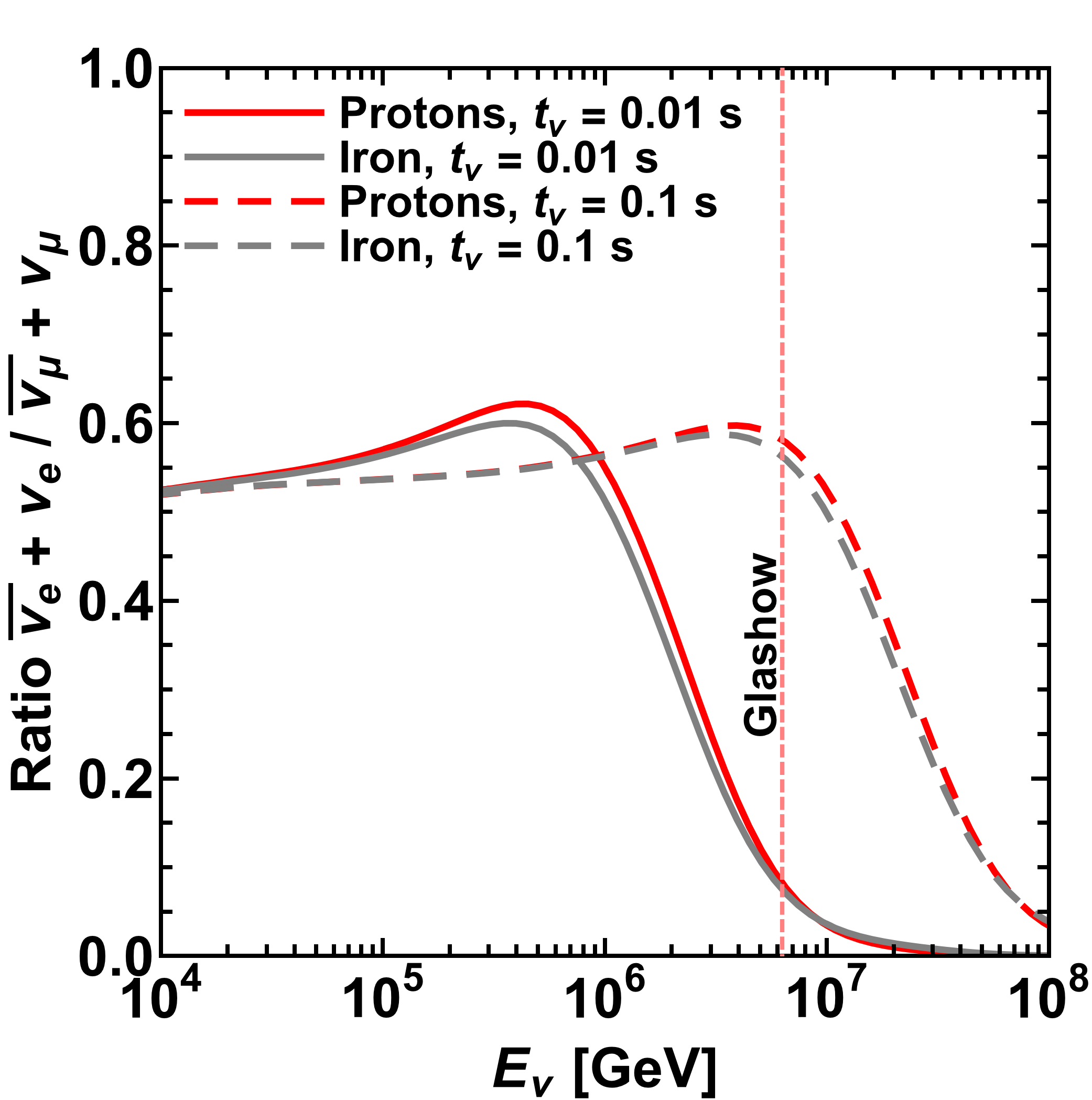}
\caption{\textit{Left panel: expected number of Glashow events as a function of exposure in the GRB case ($L=10^{52}$ erg/s, $t_v = 0.01$ s) for different isotopes with muon damping. 
Right panel: flavor ratio of electron neutrinos to muon neutrinos (without flavor mixing) as a function of the energy showing the transition to the muon damped regime.}}
\label{mudamp}
\end{figure*}

The proper lifetime of charged pions is by a factor of 85 shorter compared to muons. One can therefore construct source classes in which muons lose a large fraction of their energy (\eg\ via synchrotron cooling) before they can decay, whereas the pion flux is not yet attenuated~\cite{dampedMU,Kashti:2005qa,Lipari:2007su,Reynoso:2008gs,Hummer:2010ai,Winter:2012xq}. 
The damped $\mu^+$ mode in ${\rm p} \gamma$ interactions can occur in sources with high radiation densities and, consequently, high magnetic fields. Examples for such astrophysical source candidates are GRBs, microquasars~\cite{Reynoso:2008gs}, and tidal disruption flares~\cite{Wang:2011ip,Lunardini:2016xwi}.

In the damped $\mu^+$ mode, only $\numu$  are produced at the source by
\begin{equation}
\label{eq:damped-pi-plus}			
\pi^+\ra\mu^+ \numu \, .
\end{equation}
This means that  no antineutrinos are produced at all, including no $\nuebar$'s. 
In the presence of a $\pi^-$ contamination, one has 
\begin{equation}
\label{eq:damped-pi-minus}
\pi^-\ra\mu^- \numubar  \stackrel{\rm mix}{\lra} \xi_{\nuebar}^f = \frac{2}{9} 
\end{equation}
and therefore a potentially large effect on the  rate of the Glashow events in the damped muon case.

We illustrate the impact of the muon damping in Fig.~\ref{mudamp}, right panel, for two different parameter sets for protons and iron.\footnote{In the GRB example in this work, the muon damping is actually limited by adiabatic cooling. This is the reason why $t_v$, to which the  adiabatic cooling rate is inversely proportional, is relevant for the  critical energy above which muon damping is important -- and it is not just sufficient to decrease the magnetic loading (which is assumed to be in equipartition with the photon energy).} In this scenario, it is important to evaluate the amount of $\pi^-$, because if there are no $\pi^-$, no $\bar{\nu}$ will be produced and the signal expected at Glashow resonance would be null. Anyway, in the above sections we pointed out that ${\rm p}\gamma$ without $\pi^-$ is not a realistic case, so also in the muon damped case there will be a significant fraction of $\nuebar$ after flavor mixing. In Fig. \ref{mudamp}, we can see a comparison between the expected signal for protons and iron in a muon damped scenario. For protons the situation is not so different from the ideal ${\rm p}\gamma$ scenario, whereas for iron the situation is more similar to the pp scenario, due to the intrinsic presence of neutrons (\ie, $\pi^-$) at the source. 
Very interestingly, the relative distance between ${\rm A}\gamma$ and ${\rm p}\gamma$ hardly changes compared to the full pion decay chain, and about 115 times the yearly IC86 exposure is required to discriminate ${\rm A}\gamma$ from ${\rm p} \gamma$. In that case, one would however need additional observables (such as the flavor composition) to establish the muon damping (which is plausible for IceCube-Gen2, see Fig.~3 in \Ref~\cite{Bustamante:2015waa}).

\subsection{Scenario discrimination}

\begin{table*}[t]
\caption{\textit{IC86 equivalent exposure required to distinguish between different scenarios at the 90\% C.L., considering two spectral indices of the detected astrophysical neutrino flux. If the required exposure is greater than few hundred years, the symbol ``$\infty$'' is used. IceCube-Gen2 potentially accessible exposures are in bold (less than 120 equivalent years). With ``Damped $\mu$'', we denote the proton damped $\mu$ scenario, whereas with ``optically thick'', we refer to the $p\gamma$ scenario (protons, not heavy nuclei) with $L=10^{53}$ erg/s. Note that the $p \gamma$ scenario is slightly different from the ``Protons'' scenario used in \figu{grb}.}}

\begin{center}
\begin{tabular}{|c|cccccc|}
\hline
 Data $\rightarrow$  & \multicolumn{6}{c|}{$\alpha=2$} \\
Theory $\downarrow$ 
& Ideal ${\rm p}\gamma$ &  ${\rm p}\gamma$ & pp & ${\rm Fe} \gamma$ & Damped $\mu$ & Opt. thick \\
\hline
Ideal ${\rm p}\gamma$ &  $\cdot$  & \bf{81} & \bf{32} & \bf{26} & $\infty$ & \bf{55} \\
${\rm p}\gamma$ & \bf{53} & $\cdot$ & 156 & \bf{93} & 144 & $\infty$ \\
pp  & \bf{16}  & 121 &   $\cdot$ & $\infty$  & \bf{28}  & $\infty$ \\
${\rm Fe} \gamma$ & \bf{12}  & \bf{66} & $\infty$ & $\cdot$ & \bf{20}  & 122 \\
Damped $\mu$  & $\infty$  & $\infty$ & \bf{48} & \bf{37} & $\cdot$ & \bf{104}\\
Opt. thick & \bf{33} & $\infty$ & $\infty$ & 162 &  \bf{72} & $\cdot$ \\
\hline
\end{tabular} \hspace*{0.2cm} 
\begin{tabular}{|c|cccccc|}
\hline

 Data $\rightarrow$  & \multicolumn{6}{c|}{$\alpha=2.50$} \\
Theory $\downarrow$ 
& Ideal ${\rm p}\gamma$ &  ${\rm p}\gamma$ & pp & ${\rm Fe} \gamma$ & Damped $\mu$ & Opt. thick \\
\hline
Ideal ${\rm p}\gamma$ &  $\cdot$  & $\infty$ & \bf{81} & \bf{67} & $\infty$ & 140 \\
${\rm p}\gamma$ & 130 & $\cdot$ &$\infty$ & $\infty$ & $\infty$ & $\infty$  \\
pp  & \bf{39}  & $\infty$ &   $\cdot$ & $\infty$  & \bf{70}  & $\infty$ \\
${\rm Fe} \gamma$ & \bf{29}  & 168 & $\infty$ & $\cdot$ & \bf{49}  &  $\infty$\\
Damped $\mu$  & $\infty$  & $\infty$ & 123 & \bf{95} & $\cdot$  & $\infty$  \\
Opt. thick & \bf{81} & $\infty$ & $\infty$ & $\infty$ &182 & $\cdot$ \\
\hline
\end{tabular}
\end{center}
\label{tab:esp2}
\end{table*}

We summarize the discrimination power among different scenarios in terms of the IC86 equivalent exposure in \Tab~\ref{tab:esp2} for two spectral indices of the observed diffuse flux. This table illustrates that (i) IC86 has absolutely no chance to use the Glashow resonance for source diagnostics within ten years, (ii) IceCube-Gen2 can potentially discriminate many scenarios if the observed neutrino spectrum is hard enough, and (iii) this result depends very much on the spectral index of the observed spectrum (compare number of bold entries in left and right panels). Note again that there is a slight asymmetry between theory and data. The table also suggests that discrimination between the conventional pp and ${\rm p}\gamma$ scenarios including realistic assumptions for the source will be extremely challenging, even for IceCube-Gen2, while the discrimination between the ${\rm p}\gamma$ and $A \gamma$ production mechanisms may actually be the most interesting application (note that the ${\rm p}\gamma$ scenario here is slightly different from \figu{grb}).

\section{Interpretation of non-observation of Glashow events}
\label{sec:non}

\begin{table*}[t]
\caption{\textit{Equivalent exposure of IC86 required to observe more than 0 events with a 90\% of confidence level. }}
\begin{center}
\begin{tabular}{|c|c|cccccc|}
\hline
Dataset  &Spectral index & Ideal ${\rm p}\gamma$   & ${\rm p}\gamma$ & pp & ${\rm Fe} \gamma$ & Damped $\mu$  & Optically thick   \\
\hline
Global fit \cite{Aartsen:2015knd}& $\alpha = 2.5$ & 33  & 22 & 18 & 16 & 32 & 20 \\
Through-going muons \cite{Aartsen:2015zva,Aartsen:2016xlq}  & $\alpha \simeq 2$ & 15 & 9.7 & 7.6 & 7.1 & 12  &8.9  \\
\hline
\end{tabular}
\end{center}
\label{esp1}
\end{table*}

It is a frequently asked question what the non-observation of Glashow events actually means, and when conventional scenarios get under tension. From our figures one can extract the number of years needed to exclude zero events by looking for the $T_{\rm exp}$ value where the low edge of the shaded region exceeds zero events. This approach would be roughly consistent with the Feldman-Cousins approach~\cite{Feldman:1997qc}, where 2.44 events are needed to obtain a 90\% C.L.\ exclusion limit for the theory in case of of zero detected events with zero background (here the role of theory and data is exchanged). 

We show in \Tab~\ref{esp1} the required IC86 equivalent exposure of IC86 to exclude zero events at the 90\% confidence level.  In the table different spectral indices and normalizations (Global fit and Through-going muons) are used. We find that the iron and pp scenario should produce a non-zero event rate under the hypothesis of a hard spectrum, whereas the non-observation of events in IC86 becomes not indicative if the true spectrum is softer. Very important information will therefore come from the better determination of the diffuse spectrum. The only case which IC86 alone can possibly exclude are heavier nuclei primaries for $A \gamma$ interactions within about 16 years of operation even in the worst case assumption for the spectral index.

\section{Summary and conclusions}
\label{sec:conc}

We have discussed the potential of Glashow resonant events, sensitive to $\nuebar$ at the detector, as a discriminator among generic scenarios for the astrophysical neutrino production. We have included realistic flavor mixing (and its uncertainties),  the kinematics of the secondary (muon, pion) decays, and used Monte Carlo event generators to model pp ({\sc Sibyll} 2.3, {\sc Epos-lhc} and {\sc Qgsjet}-II-04) and ${\rm p}\gamma$ ({\sc Sophia}) interactions. We have also studied the optically thick (to photohadronic interactions) case, $A \gamma$ interactions for neutrino production from nuclei, and damped muon sources.

The often used reference assumption of equal production of $\pi^+$ and $\pi^-$ for pp sources only holds if the spectral index $\simeq 2$, whereas softer spectra lead to significant deviations of the electron antineutrino faction of up to 30\%. For ${\rm p} \gamma$ interactions, it is well known that multi-pion and other processes lead to a contamination of $\pi^-$, leading to sources more similar to pp than the reference production scenario using the $\Delta$-resonance approximation. We have demonstrated that the discrimination between the pp and ${\rm p}\gamma$ 
mechanisms in the source becomes extremely challenging, even for IceCube-Gen2 and hard spectral indices, if realistic assumptions for the neutrino production are included.

As far as  different source conditions are concerned, optically thick (to photohadronic interactions) sources will lead to an enhancement of neutrons in the source, which will prohibit any discrimination power from pp. If heavier (than protons) nuclei are the neutrino primaries, photohadronic $A \gamma$ interactions will produce similar results to pp interactions with even somewhat larger Glashow rates, as nuclei typically carry more neutrons than protons, enhancing the $\pi^-$ production.  Very interestingly, the Glashow resonance  can be used to discriminate between ${\rm p} \gamma$ and $A \gamma$ interactions, and may therefore be the smoking gun signature for heavier nuclei in the sources. 
This interpretation has to rely on the assumption that the radiation density in the sources, typically known for specific source classes (such as AGNs or GRBs) must be high enough such that photohadronic interactions dominate over pp or Ap interactions.

The discrimination power among different scenarios is addressed in  \Tab~\ref{tab:esp2}, which clearly depends on the spectral index of the diffuse neutrino flux. For $\alpha \lesssim 2$, many scenarios can be discriminated against each other with IceCube-Gen2, while for $\alpha \gtrsim 2.5$ only few scenarios can be discriminated. Finally, we have demonstrated that the non-observation of Glashow resonance events will exert pressure on almost all (including ${\rm p}\gamma$) scenarios if $\alpha \lesssim 2$, which means that the observation of at least one Glashow resonant event is likely after 10 years of operation of IC86 if the spectrum turns out to be hard (or there is a spectral hardening). A non-observation in that case may be, for example, interpreted as the effect of muon damping, and may be indicative for a source class with strong magnetic fields.

We have also demonstrated that the only case in which the Glashow events rate is expected in IC86 within about 16 years -- almost independent of the spectral index -- is the neutrino production off nuclei. Therefore, the Glashow resonance is a uniquely different way to probe the cosmic ray composition. For example, the non-observation of Glashow events in IceCube contradicts a primary composition of pure iron at an energy of about $7 \, \mathrm{EeV}$, meaning that a possible iron population of UHECRs at that energy would come from different sources than the neutrinos. This result hardly depends on the cosmic ray composition for nuclei heavier than protons, as neutrons in the nuclei are much more efficient in production Glashow resonant events -- as long as the neutrons do not have time to decay (which would enhance the protons compared to the neutrons again).
The neutrino diagnostics for neutrino production off heavy nuclei may therefore be the key application of the Glashow resonance.

We conclude that reasonable applications of the Glashow resonance presumably lie beyond existing technology. However, we have demonstrated that IceCube-Gen2 potentially has enough exposure to discriminate among different scenarios. For photohadronic interactions, the Glashow resonance may be a unique way to discriminate protons from heavier nuclei in the sources.

\acknowledgments 

 This project has received funding from the European Research Council (ERC) under the European Union’s Horizon 2020 research and innovation programme (Grant No. 646623). The research of TJW was partially supported by DoE grant No.\ DE-SC0011981. TJW and WW acknowledge support from the KITP Santa Barbara for the program ``Present and future neutrino physics'', where first discussions on this topic started. 

\appendix

\section{Neutrinos from hadronic interactions}
\label{app:zeta}

A detailed modeling of hadronic interactions is possible in a semi-analytical approach that has been widely used for classical calculations of atmospheric lepton fluxes \cite{gaisser_book2}. The approach is based on spectrum weighted moments, so called $Z$ factors, that represent evaluated source functions in approximate solutions of the coupled cascade equations for power-law spectra.

\subsection{Pion production in hadronic interactions}
\label{zpion}

A realistic description of inclusive pion production in sources with power-law projectile spectra can be obtained through the $Z$ factors 
\begin{equation}
\label{equ:zeta_factor}
Z_{{\rm p}\pi^+}(\alpha) =\int_0^1 x^{\alpha-1} \ 
	\frac{\rmd  N({\rm pp} \to {\pi^+} + {\rm anything})}{\rmd  x} \ \rmd x
\end{equation}
where $-\alpha$ ($\alpha>0$) is the differential spectral index of the projectiles (protons), $x = E_{\pi^+} / E_p$ and $\rmd  N_{\pi^+}/\rmd  x$ is the inclusive pion spectrum that can be obtained from Monte Carlo event generators or accelerator data. 
\begin{table}[ht!]
\caption{\textit{Spectrum weighted moments of pions in pp interactions.}}
\label{zeta2}
\begin{center}
\begin{tabular}{|c|cccc|}
\hline
$\alpha$ & $Z_{{\rm p}\pi^+}$ & $Z_{{\rm p}\pi^-}$ & $Z_{{\rm p}\pi^0}$ &$N_{\pi^+}/N_{\pi^-}$ \\
\hline
$2.0$ &$ 0.157 \pm 0.014$ & $0.126 \pm 0.017$ & $ 0.142 \pm 0.003$  & $1.25 \pm 0.20$\\
$2.3$ & $0.076 \pm 0.007$ &$ 0.057 \pm 0.004$ & $0.054 \pm 0.016$ & $1.34 \pm 0.15$ \\
$2.6$ &$0.042 \pm 0.006 $&$0.029 \pm 0.002$ & $0.031 \pm 0.001 $&$1.43 \pm 0.24$ \\
\hline
\end{tabular}
\end{center}
\end{table} 
Table~\ref{zeta2} lists $Z$ factors of the pions from pp interaction; they are obtained from running the hadronic interaction models {\sc Sibyll} 2.3 \cite{Ahn:2009wx,Riehn:2015tbd}, {\sc Epos-lhc} \cite{Pierog:2013ria} and {\sc Qgsjet}-II-04 \cite{Ostapchenko:2010vb}, where the target proton is assumed to be at rest, while the projectile has an energy of 100 PeV (remember that the mean energy of a neutrino is about 5\% of the energy of the proton, so this is the right energy to consider for the Glashow resonance). The values shown in the table are the mean values obtained with the three models and the uncertainties are the standard deviations. A realistic estimate of the pion charge ratio is simply the ratio of the corresponding $Z$ factors. As listed in \Tab~\ref{zeta2}, the charge ratio $\pi^+/\pi^-$ can be very different ($\sim$ 25\%-45\%) from the ideal condition $\pi^+ \simeq \pi^- \simeq \pi^0$, that is often used in the treatment of pp interactions. It is important to take into account this effect in order to make a precise estimation of the expected rate of Glashow events. 

\subsection{Neutrinos from pion decay}
\label{zetanu}

The energy distributions of neutrinos from pion decay $\rmd N(\pi \to \mu \nu)/\rmd x$ are reported in \Ref~\cite{Kelner:2006tc} and analytically obtained. Analogously to the approach in the previous section, one can define the decay $Z$ factor (see \Ref~\cite{Lipari:1993hd}, also for a comparison of the values), \eg\
\begin{equation}
\label{equ:decay_Z_factor}
Z_{\pi\numu^1}(\alpha)=\int_0^1 x^{\alpha-1} \ \frac{\rmd N(\pi \to \mu \numu)}{\rmd x} \  dx.
\end{equation}
Branching ratios are included in the energy distributions. For more complex production channels one has to sum over the distributions of each decay channel times the branching ratio. The values obtained for different spectral indices are reported in \Tab~\ref{zeta}.
\begin{table}[h]
\caption{\textit{$Z$ factors of neutrinos from pion decay. $Z_{\pi\numu^1}$ is the muon (anti)neutrino from $\pi$ decay. $Z_{\pi\numu^2}$ is the muon (anti)neutrino and $Z_{\pi \nue}$ the electron (anti)neutrino from the subsequent $\mu$ decay.}}
\label{zeta}
\begin{center}
\begin{tabular}{|c|ccc|}
\hline
 $\alpha$ & $Z_{\pi\numu^1}$ & $Z_{\pi\numu^2}$ & $Z_{\pi \nue}$ \\
\hline
$2.0$ & 0.213 & 0.265 & 0.257 \\
$2.3$ & 0.144 & 0.194 & 0.188 \\
$2.6$ &0.098 &0.146 & 0.141 \\
\hline
\end{tabular}
\end{center}
\end{table} 
Using the $Z$ factor approach, one obtains the neutrino fluxes at the source from
\begin{equation}
\begin{split}
\phi_{_e} &= \phi_p \times  Z_{{\rm p}\pi^+}  \times Z_{\pi\nue}, \\
\phi_{\bar e} &= \phi_p \times Z_{{\rm p}\pi^-} \times Z_{\pi\nue}, \\
\phi_{\mu} &= \phi_p \times (Z_{{\rm p}\pi^+} \times Z_{\pi\numu^1}+Z_{{\rm p}\pi^-} \times Z_{\pi\numu^2}), \\
\phi_{\numubar} &= \phi_p \times (Z_{{\rm p}\pi^+}  \times Z_{\pi\numu^2}+Z_{{\rm p}\pi^-} \times Z_{\pi\numu^1}),
\end{split}
\label{fracs2}
\end{equation}
where $\phi_p$ is the flux of protons, $\phi_{\ell}$ is the flux of neutrinos of flavor $\ell$, $Z_{\pi\numu^1}$ is the muon neutrino (antineutrino) from $\pi$ decay. $Z_{\pi\numu^2}$ is the muon neutrino (antineutrino) and $Z_{\pi \nue}$ the electron neutrino (antineutrino) from the subsequent $\mu$ decay. Note that this solution assumes that pions do not (re-)interact and that all pions and muons decay. Additional pion production by secondary neutrons is verified to be negligible for the calculation of flavor ratios. Let us remark that in the reference pp scenario $Z_{{\rm p}\pi^+} = Z_{{\rm p}\pi^-}$, whereas in the reference ${\rm p}\gamma$ scenario $Z_{{\rm p}\pi^-}=0$; in the reference scenario it is therefore not important to know the exact value of $Z_{{\rm p}\pi^+}$ because it cancels when the flavor ratio is calculated.

\begin{table}[h!]
\caption{\textit{Flavor composition at the source for a realistic pp scenario and different spectral indices $\alpha$. The value $\delta$ represents the maximum deviation between the amount of neutrinos of a certain flavor and the simplified flavor composition usually assumed, i.e. }($\frac{1}{6}:\frac{2}{6}:0:\frac{1}{6}:\frac{2}{6}:0$).}
\label{flvcomp}
\begin{center}
\begin{tabular}{|c|c|ccccccc|}
\hline
Source & $\alpha$ & $\xi_{\nue}$ & $\xi_{\numu}$ & $\xi_{\nutau}$ & $\xi_{\nuebar}$ & $\xi_{\numubar}$ & $\xi_{\nutaubar}$ & $\delta$\\
\hline
pp & 2.0 & 0.194 & 0.321 & 0 & 0.156 & 0.329 & 0 & 16\%  \\
pp & 2.3 & 0.204 & 0.314 & 0 & 0.153 & 0.329 & 0 & 22\% \\
pp & 2.6 & 0.217 & 0.305 & 0 & 0.149 & 0.329 & 0  & 30\% \\
\hline
\end{tabular}
\end{center}
\end{table} 
In \Tab~\ref{flvcomp} the flavor composition from a realistic pp scenario is reported; the difference between the simplified flavor composition usually adopted can reach 30\% with a high spectral index.

\section{The effect of Kaons in the  \texorpdfstring{pp}{pp} mechanism}
\label{app:kmesons}

In this section the effect of kaons is analyzed, showing that it is a second order effect that can be neglected. A general discussion relative to the contribution of kaons in $pp$ interaction, not limited to the power law hypothesis, is done in \cite{vilvis2008}. 

In the case of power law spectrum of primary protons and $pp$ interaction,
the contribution of kaons can be added using the following formulae:
\begin{equation}
\begin{split}
\phi_{_e} = \phi_p \times [&Z_{{\rm p}\pi^+}  \times Z_{\pi\nue} + Z_{{\rm p} {\rm K}^+}(Z_{{\rm K}\nue}+Z_{{\rm K}\mu\nue}) ] \\
\phi_{\bar e}= \phi_p \times [ &Z_{{\rm p}\pi^-}  \times Z_{\pi\nue} + Z_{{\rm p} {\rm K}^-}(Z_{{\rm K} \nue}+Z_{{\rm K}\mu\nue})] \\
\phi_{\mu}=\phi_p \times [ &Z_{{\rm p}\pi^+} \times Z_{\pi\numu^1}+Z_{{\rm p}\pi^-} \times Z_{\pi\numu^2}\\ &+ Z_{{\rm p} {\rm K}^+} \times Z_{{\rm K} \numu} + Z_{{\rm p} {\rm K}^-} \times Z_{{\rm K} \mu \numu} ] \\
\phi_{\bar \mu}=\phi_p \times [&Z_{{\rm p}\pi^+} \times Z_{\pi\numu^2}+Z_{{\rm p}\pi^-}\times Z_{\pi\numu^1} \\& + Z_{{\rm p} {\rm K}^-} \times Z_{{\rm K} \numu} + Z_{{\rm p} {\rm K}^+} \times Z_{{\rm K} \mu \numu}  ].
\end{split}
\end{equation} 
The factors $Z_{{\rm p}K}$ for $\alpha=2.0, \ 2.3, \ 2.6$ are 
\begin{equation}
\begin{split}
Z_{{\rm p} {\rm K}^+}&=(0.028,0.014,0.009) \\ 
Z_{{\rm p} {\rm K}^-}&=(0.018,0.008,0.004).
\end{split}
\end{equation}
They are obtained by averaging over the values from {\sc Sibyll} 2.3, {\sc Epos-lhc} and {\sc Qgsjet}-II-04; an asymmetry between the number of ${\rm K}^+$ and ${\rm K}^-$ is present, as in the pions. 
The $Z_{{\rm K}\nu_\ell}$ and $Z_{{\rm K}\mu\nu_\ell}$ are given in \Ref~\cite{Lipari:1993hd} for $\alpha=2.0,\ 2.7, \ 3.7$ and the values for $\alpha=2.3$ and $\alpha=2.6$ were found using a linear interpolation. 

The flavor composition hardly changes for the ${\rm p}\gamma$ case, in which we assume that no kaons are produced. This assumption is justified because the ${\rm p}\gamma$ process happens near the energy threshold of $\Delta^+$ production. In the case of  pp interactions, the flavor composition slightly changes, as reported in Table \ref{flvcompk}. Adding kaons increases the difference between the often used standard composition up to the level of 35\%. Anyway, the main contribution is given by the asymmetry between the number of $\pi^+$ and the number of $\pi^-$, that are not produced in equal number as usually assumed. The contribution of kaons is less relevant than the contributions given by $Z$ factors $Z_{{\rm p}\pi}$ and $Z_{\pi\nu}$. We find that from the \Tab~\ref{flvcompk} by looking at $\delta_K$ that represents the deviation from the Monte Carlo cases of \Tab~\ref{flvcomp} including the contribution of kaons.

\begin{table}[t!]
\caption{\textit{Flavor composition at the source in the pp scenario scenario including the contribution from kaons. Here $\delta_K$ represents the deviation from the Monte Carlo cases of \Tab~\ref{flvcomp}.
}}
\label{flvcompk}
\begin{center}
\begin{tabular}{|c|c|ccccccc|}
\hline
Source & $\alpha$ & $\xi_{\nue}$ & $\xi_{\numu}$ & $\xi_{\nutau}$ & ${\xi}_{\nuebar}$ & ${\xi}_{\numubar}$ & ${\xi}_{\nutaubar}$ & $\delta_K$ \\
\hline
pp & 2.0 & 0.206 & 0.320 & 0 & 0.159 & 0.315 & 0 & 6\%   \\
pp & 2.3 & 0.216 & 0.321 & 0 & 0.152 & 0.311 & 0 & 5\%  \\
pp & 2.6 & 0.226 & 0.326 & 0 & 0.142 & 0.306 & 0  & 3\% \\
\hline
\end{tabular}
\end{center}
\end{table} 

\section{Cross check with IceCube effective areas} 
\label{app:crosscheck}

A useful cross check of our calculation can be done by directly using the IceCube effective areas for the evaluation of the expected number of resonant events. The effective areas are provided for each flavor as averages of neutrino and antineutrino contributions.

The effective area of $\nutau$ ($A_{\nutau}$) is simply related to that of the deep inelastic scattering $A_{\mbox{\tiny DIS}}$, 
$$
A_{\nutau}=A_{\mbox{\tiny DIS}}
$$
On the other side, the effective area of $\nue$ ($A_{\nue}$) is the sum of two contributions, one from deep inelastic scattering and one from Glashow resonance ($A_{\mbox{\tiny G}}$), following the relation: 
$$
A_{\nue}=A_{\mbox{\tiny DIS}} + \frac{1}{2} A_{\mbox{\tiny G}}
$$ 
As remarked in \Ref~\cite{Palladino:2015uoa} and in Fig. 7 of \Ref~\cite{Aartsen:2013jdh} the effective area of $\nue$ \textit{contains an assumption on the fluxes of $\nue$ and $\nuebar$}; particularly it is considered that the flux of $\nue$ and $\nuebar$ is the same after oscillations. This justifies the factor $\frac{1}{2}$, since only $\nuebar$ interacts via Glashow resonance. The only way to give effective areas that do not depend on the production mechanism is to provide four different effective areas (for $\nue, \ \nuebar, \ \numu, \ \nutau$).
Therefore the Glashow resonance effective area is approximately given by
\begin{equation}
A_{\mbox{\tiny G}}(E_\nu)=2 \ [A_{\nue}(E_\nu)-A_{\nutau}(E_\nu)],
\end{equation}
because at high energies the effective areas of $\nue$ and $\nutau$ relative to the deep inelastic scattering are in good approximation equal each other. 
To evaluate the number of resonant events $N$, we can use the general formula
\begin{equation}
N^{\mbox{\tiny G}}=4\pi \ T \  \int_0^\infty \xi^f_{\nuebar} \ \phi^f_{\mbox{\tiny 3f}}  \times A_{\mbox{\tiny G}}(E_\nu) \ \rmd E_\nu \, ,
\end{equation} 
where $\xi^f_{\nuebar}$ depends on the production mechanism. 
This is the method given by the IceCube collaboration in \Ref~\cite{Aartsen:2013jdh} to correctly use the effective areas. 
Using the best fit value of the Through-going muon analysis, the number of events per year is equal to 0.52. The difference from the 0.35 events that we found for the pp mechanism is expected, since we only take into account the events that give rise to a signal around 6.3 PeV. Further, the effective areas do not contain information on the deposited energy from the neutrino interaction process, so it is only possible to obtain the total number of events produced at the Glashow resonance energy. However, about 34\% of the events can deposit energy in the range between 0 and 6.3 PeV and it is impossible to distinguish those events from normal cascades by deep inelastic scattering. Only 64\% of the events can deposit all energy as hadronic shower (see \Ref~\cite{Palladino:2015uoa} for a detailed discussion). For this reason the above number must be multiplied by the branching ratio of $W^- \rightarrow \ \mbox{hadrons}$, resulting in our value of $\sim 0.35$ events per year.

 \bibliographystyle{apsrev4-1}
%

\end{document}